\documentclass[a4paper,fleqn,usenatbib]{mnras}

\usepackage{etoolbox}
\makeatletter
\patchcmd\@combinedblfloats{\box\@outputbox}{\unvbox\@outputbox}{}{%
   \errmessage{\noexpand\@combinedblfloats could not be patched}%
}%

\usepackage{Times}

\usepackage[T1]{fontenc}
\usepackage{ae,aecompl}

\usepackage{graphicx}	
\usepackage{amsmath}	
\usepackage{amssymb}	
\usepackage{tablefootnote}
\usepackage{gensymb}

\title[VLT/MUSE observations of VR7]{Resolved Lyman-$\alpha$ properties of a luminous Lyman-break galaxy in a large ionised bubble at $z=6.53$ }

\author[J. Matthee et al.]{Jorryt Matthee$^{1}$\thanks{Zwicky Fellow -- mattheej@phys.ethz.ch}, David Sobral$^{2}$, Max Gronke$^{3}$\thanks{Hubble Fellow}, Gabriele Pezzulli$^{1}$, \newauthor 
Sebastiano Cantalupo$^{1}$, Huub R\"ottgering$^{4}$, Behnam Darvish$^{5}$,  S\'{e}rgio Santos$^{2}$  \\
$^{1}$ Department of Physiscs, ETH Z\"urich, Wolfgang-Pauli-Strasse 27, 8093 Z\"urich, Switzerland\\
$^{2}$ Department of Physics, Lancaster University, Lancaster, LA1 4YB, UK\\
$^{3}$ Department of Physics and Astronomy, University of California, Santa Barbara, USA\\
$^{4}$ Leiden Observatory, Leiden University, PO\ Box 9513, NL-2300 RA Leiden, The Netherlands\\
$^{5}$ Cahill Center for Astrophysics, California Institute of Technology, 1216 East California Boulevard, Pasadena, CA 91125, USA \\
}

\pubyear{2019}

\begin{document}
\label{firstpage}
\pagerange{\pageref{firstpage}--\pageref{lastpage}}
\maketitle

\begin{abstract}
The observed properties of the Lyman-$\alpha$ (Ly$\alpha$) emission line are a powerful probe of neutral gas in and around galaxies. We present spatially resolved Ly$\alpha$ spectroscopy with VLT/MUSE targeting VR7, a UV-luminous galaxy at $z=6.532$ with moderate Ly$\alpha$ equivalent width (EW$_0\approx38$ {\AA}). These data are combined with deep resolved [CII]$_{\rm 158 \mu m}$ spectroscopy obtained with ALMA and UV imaging from {\it HST} and we also detect UV continuum with MUSE. Ly$\alpha$ emission is clearly detected with S/N $\approx40$ and FWHM of 374 km s$^{-1}$. Ly$\alpha$ and [CII] are similarly extended beyond the UV, with effective radius r$_{\rm eff} = 2.1\pm0.2$ kpc for a single exponential model or r$_{\rm eff, Ly\alpha, halo} = 3.45^{+1.08}_{-0.87}$ kpc when measured jointly with the UV continuum. The Ly$\alpha$ profile is broader and redshifted with respect to the [CII] line (by 213 km s$^{-1}$), but there are spatial variations that are qualitatively similar in both lines and coincide with resolved UV components. This suggests that the emission originates from two components with plausibly different HI column densities. We place VR7 in the context of other galaxies at similar and lower redshift. The Ly$\alpha$ halo scale length is similar at different redshifts and velocity shifts with respect to the systemic are typically smaller. Overall, we find little indications of a more neutral vicinity at higher redshift. This means that the local ($\sim 10$ kpc) neutral gas conditions that determine the observed Ly$\alpha$ properties in VR7 resemble the conditions in post-re-ionisation galaxies.
\end{abstract} 



\section{Introduction}
The re-ionisation epoch marks the transition of intergalactic hydrogen from neutral to ionised. Such transition is thought to occur between $z\approx6-10$ \citep{Fan2006,Planck2015,Banados2018}. However, the exact timing and topology of re-ionisation and the major origin of ionising photons are not well known.

Due to the sensitivity of the Lyman-$\alpha$ (Ly$\alpha$; $\lambda_0=1215.67$ {\AA}) equivalent width (EW) in high-redshift galaxies to neutral hydrogen \citep[e.g.][]{Dijkstra2007}, it has been extensively explored as a probe of the evolving neutral fraction in the epoch of re-ionisation \citep[e.g.][]{Stark2010,Pentericci2014}. However, in addition to the EW, more subtle variations in the Ly$\alpha$ line profile and spatial extent are also expected \citep[e.g.][]{MasRibas2017}, but are difficult to explore at $z>6$ \citep[c.f.][]{Hu2010,Momose2014,Kikuma2019}.

One effect of an increasing neutral fraction is a reduced transmission of Ly$\alpha$ photons at increasingly redder wavelengths with respect to the systemic redshift \citep[e.g.][]{Laursen2011}. How this affects the observed Ly$\alpha$ properties depends on the velocity shift between Ly$\alpha$ and the systemic redshift \citep[e.g.][]{Choudhury2014,Mason2018}, for example due to outflows \citep[e.g.][]{Erb2014,RiveraThorsen2015}. 

Currently, velocity shifts between the peak of the Ly$\alpha$ line and the systemic redshift at $z>6$ are measured with other rest-frame UV lines \citep[e.g.][]{Stark2017} or through measurements of far-infrared lines with ALMA \citep[e.g.][]{Pentericci2016,Hashimoto2018Dragons}. The interpretation of {\it observed} velocity shifts at $z>6$ is challenging: a large observed shift could be intrinsic if, for example, outflows redshift Ly$\alpha$ photons out of the resonance wavelength \citep{Verhamme2006} before encountering significant amount of neutral hydrogen so that the intervening intergalactic medium (IGM) is effectively transparent \citep{DijkstraReview}. A larger observed shift could also be the consequence of a decreased transmission due to large amounts of neutral hydrogen around galaxies which in practice absorbs the bluer part of the line \citep[e.g.][]{Laursen2011,Smith2019}. Finally, large shifts could also be consequences of large velocity offsets between (blended) merging components, with lines corresponding to different components.

The spatial extent of Ly$\alpha$ emission may increase in the epoch of re-ionisation due to an increased importance of resonant scattering in the presence of more neutral hydrogen in the circum galactic medium (CGM) of galaxies. An indication of an increase in the Ly$\alpha$ scale length is found between $z=5.7$ and $z=6.6$ by \cite{Momose2014}, but those results rely heavily on stacking (see also \citealt{Santos2016}). Recently, the Multi Unit Spectroscopic Explorer (MUSE) instrument on the VLT \citep{Bacon2010} has been successful in observing extended Ly$\alpha$ emission around individual high-redshift galaxies \citep[e.g.][]{Wisotzki2015,Leclercq2017}, but so far has focused mostly on faint Lyman-$\alpha$ emitters (LAEs) at $z<6$. 

Due to their brightness, luminous Ly$\alpha$ emitters are the best targets to take studies of extended Ly$\alpha$ emission into the epoch of re-ionisation ($z>6$). Among the sample of luminous LAEs \citep[e.g.][]{Sobral2018}, VR7 is the brightest in the UV continuum (M$_{1500}=-22.4$) and consequently has a relatively typical Ly$\alpha$ equivalent width (EW), EW$_{0}=38$ {\AA} in spite of its luminous Ly$\alpha$ emission. This is similar to the typical EWs in bright UV-selected galaxies at $z\sim6$ \citep[e.g.][]{Curtis-Lake2012}. VR7 consists of two resolved components in UV and [CII] emission \citep{Matthee2019}, similarly to other luminous galaxies at $z\approx7$ \citep{Ouchi2013,Sobral2015,Matthee2017ALMA,Carniani2018Himiko,Sobral2019}. Do such multiple components influence measurements of velocity offsets with unresolved data (as e.g. \citealt{Pentericci2016})? Is the Ly$\alpha$ emission around luminous LAEs at $z=6.5$ more extended than typical LAEs at the same redshift, or more extended than galaxies at lower redshift? Can we witness any imprint of re-ionisation on the observed Ly$\alpha$ properties? These are the questions that we aim to address.

In this paper we focus on spatially resolved Ly$\alpha$ data from VLT/MUSE observations of the LAE `VR7' \citep{Matthee2017SPEC} at $z=6.532$. These data allow us to measure the Ly$\alpha$ extent and identify possible spatial variations in the Ly$\alpha$ line profile. An important aspect of this work is that we combine the resolved Ly$\alpha$ data with resolved [CII] spectroscopy from ALMA and resolved rest-frame UV imaging from {\it HST}/WFC3 \citep{Matthee2019}. This allows us to search for spatial variations in the velocity offset between Ly$\alpha$ and [CII] and to compare the extent of Ly$\alpha$ to the extent in the rest-frame UV.

The structure of this paper is as follows. We present the VLT/MUSE observations, data reduction and data quality in \S $\ref{sec:sample_observations}$. An overview of the known properties of VR7 is given in \S $\ref{sec:general}$. We investigate the environment of VR7 in \S $\ref{app:environment}$. In \S $\ref{sec:SB}$ we present the sizes and surface brightness profiles of VR7 in Ly$\alpha$, [CII] and rest-frame UV and use the rest-frame UV data to separate Ly$\alpha$ emission that is extended beyond the rest-frame UV emission. We focus on resolving the Ly$\alpha$ line profile spatially in \S $\ref{sec:linevariations}$, where we also compare it to the resolved [CII] profile. Our results are placed in context in \S $\ref{sec:discussion}$, where we discuss the Ly$\alpha$ extent and the Ly$\alpha$ velocity offset compared to other galaxies at $z\approx3-7$. We summarise our results in \S $\ref{sec:conclusions}$. Throughout the paper we use a flat $\Lambda$CDM cosmology with $\Omega_M = 0.3$, $\Omega_{\Lambda} = 0.7$ and H$_0 = 70$ km s$^{-1}$ Mpc$^{-1}$.

\begin{figure}
\includegraphics[width=8.65cm]{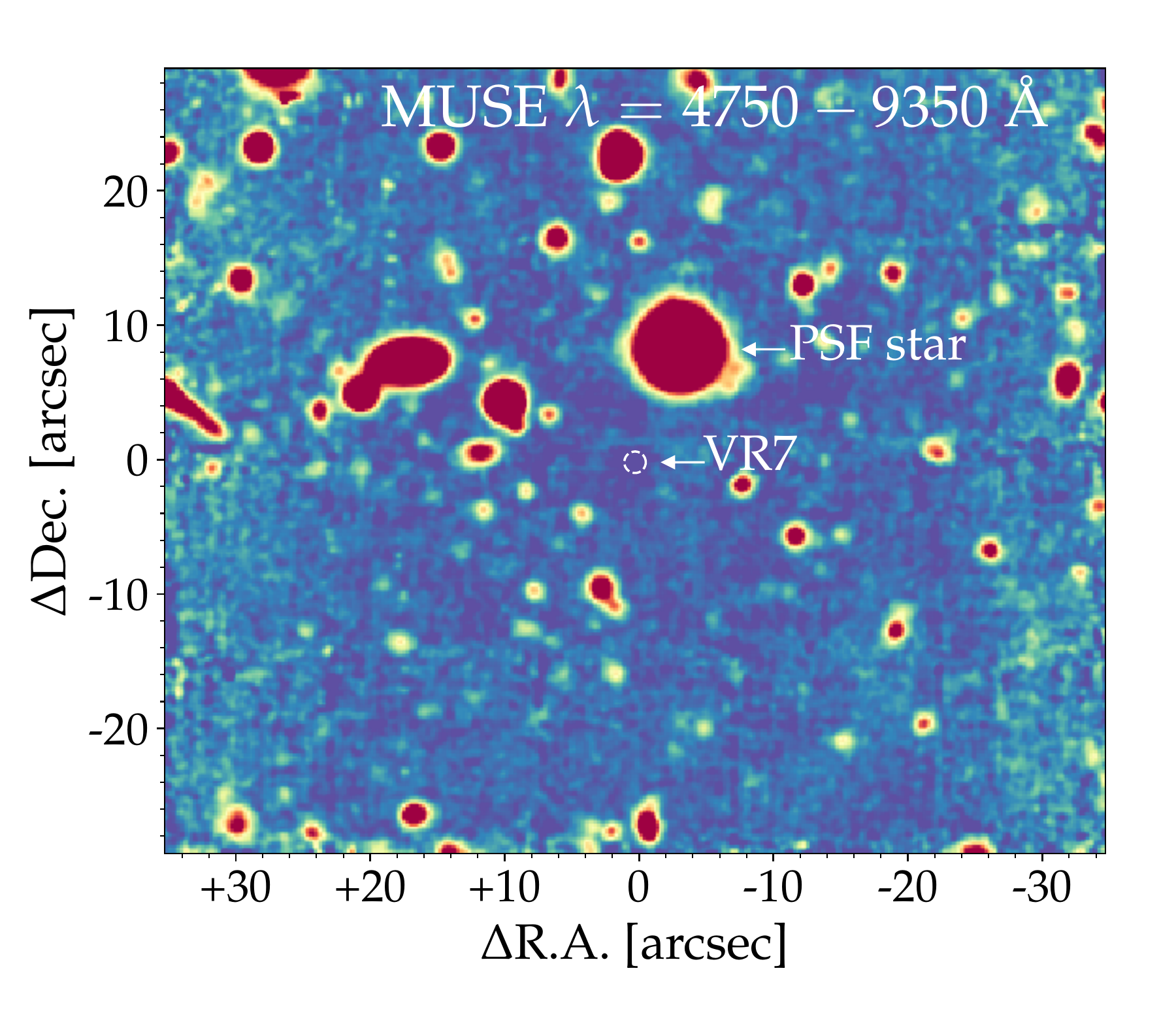}\\
\includegraphics[width=8.4cm]{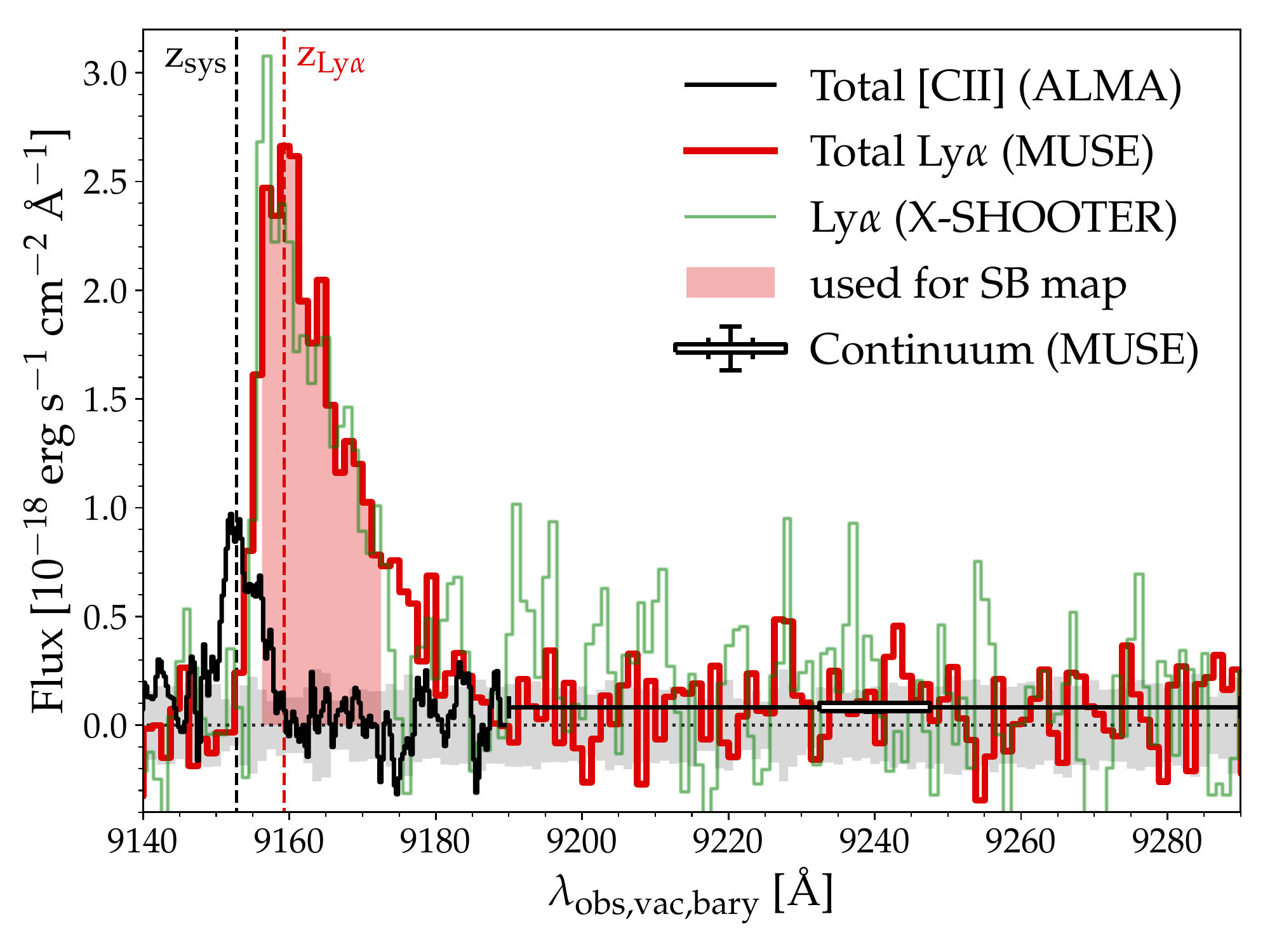}
\caption{Top: White-light ($\lambda_{\rm obs, vac}=4750-9350$ {\AA}) image of the MUSE data, where the position of VR7 (not visible in the white-light image) and the star used for PSF-estimation are highlighted. Bottom: The 1D Ly$\alpha$ spectrum of VR7 as observed by MUSE in a 1.6$''$ diameter aperture (red), the X-SHOOTER Ly$\alpha$ spectrum (\citealt{Matthee2017SPEC}; binned spectrally to match the MUSE resolution; green) and the total [CII] spectrum observed by ALMA (rescaling the observed frequency to the observed wavelength if [CII] were at $\lambda_0=1215.67$ {\AA}; black). Ly$\alpha$ emission is detected over $\approx22$ {\AA}, in 18 spectral layers and redshifted compared to [CII]. The red shaded area shows the wavelength range used to analyse the spatial distribution of Ly$\alpha$ emission ($\lambda_{\rm obs, vac}=9153.5-9170$ {\AA}; see \S $\ref{sec:SB_Lya}$). The grey shaded region shows the 1$\sigma$ noise level derived by apertures in 62 empty sky positions. The white rectangle shows the flux density in the UV continuum measured by MUSE. }
\label{fig:spectra}
\end{figure} 
 
\section{MUSE data} \label{sec:sample_observations}
\subsection{Observations \& reduction}\label{sec:spectra}
VR7 was observed with VLT/MUSE \citep{Bacon2010} in service mode through program 099.A-0462 (PI: Matthee) on 29 May, 30 June, 28 July and 21 September 2017. Observations were performed with a seeing full-width half maximum (FWHM) $\approx0.9-1.0''$ in the $V$-band and with an airmass $\approx1.1$. We used the standard Wide Field Mode with a field of view of $59.9''\times60.0''$ and pixel scale of 0.2$''$/px, with a wavelength range 480-930 nm sampled by 3681 layers of $\Delta\lambda\approx1.25$ {\AA}. Individual exposure times were 720 s. We rotated the position angle by 90 degrees after each exposure and dithered by 2$''$ in the RA/DEC direction with respect to the previous exposure. Two exposures were rotated by 120$\degree$ and 240$\degree$. Combined, our exposures uniformly fill a grid of $70''\times60''$ centred on the position of VR7 with a maximum on target exposure time of 16740 s, or 4.65 hours.

We reduce the data using the MUSE pipeline v2.2 \citep{Weilbacher2014} included in {\sc Esorex}. We first run standard prescriptions for bias subtraction, flat-fielding, illumination correction and wavelength and flux calibration for each night individually. The coordinate system of each observation is mapped to the 2MASS \citep{Skrutskie2006} reference frame in the same way as for our {\it HST} data and ground-based imaging \citep{Matthee2015,Matthee2019}. Then, we use {\sc CubEx} (Cantalupo in prep.; see \citealt{Cantalupo2019} for a description) for additional flat-fielding, the additional removal of skyline residuals and the combination of the individual exposures. {\sc CubEx} is run iteratively, where the white-light image (collapse of $\lambda=4750-9350$ {\AA}) of the first iteration has been used as a source mask for input in the final iteration. We manually add VR7 to the source mask, as it is un-detected in the white-light image due to its high redshift. 

\subsection{Quality and depth} \label{sec:quality}
We use a $I=17$ magnitude star that is 9$''$ away from the center of our pointing (see Fig. $\ref{fig:spectra}$) to measure the point spread function (PSF) of our reduced MUSE data. At the wavelength of VR7's Ly$\alpha$ line, the PSF is well described by a Moffat profile with $\beta=2.8$ \citep[similar to][]{Bacon2017} and FWHM=0.90$''$. The relative astrometric accuracy compared to the {\it HST}/WFC3 data \citep{Matthee2019} is measured by comparing the positions of 44 objects detected in the white-light image of the MUSE data-cube. We find no significant offset with an accuracy of $0.02''$.

We estimate the noise level of our data by measuring the standard deviation of the flux measured in $0.90''$ diameter apertures in 62 empty-sky positions. These positions were carefully selected based on a PSF-matched $\chi^2$-combined detection image based on ground-based $ugriz$ and {\it HST} F110W and F160W data. The centres of the empty-sky positions are $>2''$ away from any detected object. The noise level depends on wavelength due to changes in the instrument efficiency, the sky brightness and atmospheric OH features, but it is relatively constant at rms=$1\times10^{-19}$ erg s$^{-1}$ cm$^{-2}$ {\AA}$^{-1}$ in individual layers around $\lambda_{\rm obs, air} \approx 915$ nm. 

\section{General properties of VR7} \label{sec:general}
Here we present the global measurements of VR7 from the MUSE data, summarise the other multi-wavelength properties known about the galaxy and how they compare to the galaxy population at $z\sim7$.

VR7's Ly$\alpha$ line is clearly detected in the MUSE data, with an integrated signal-to-noise S/N $\approx$ 40 in a narrow-band collapsed over $\lambda_{\rm obs, air}=9153.5-9170.0$ {\AA} ($\Delta v=540$ km s$^{-1}$; found to optimise the S/N, see Fig. $\ref{fig:spectra}$). The MUSE data also detects continuum right-wards of Ly$\alpha$ with an integrated S/N=3.8 from $920-930$ nm and an AB magnitude $25.16^{+0.40}_{-0.29}$, see Appendix $\ref{app:UVcont}$ and the white symbol in Fig. $\ref{fig:spectra}$. We fit the one-dimensional Ly$\alpha$ line profile with a skewed gaussian profile that is convolved with the line spread function (LSF) of the MUSE data (LSF-FWHM is 84 km s$^{-1}$ at the observed Ly$\alpha$ wavelength; \citealt{Bacon2017}). Following e.g. \cite{Shibuya2014,Claeyssens2019}, we parametrise the skewed gaussian profile as

\begin{equation}
    f(\lambda) =  {A}\,{\rm exp} \Big( -\frac{{\Delta v}^2}{2{(a_{{\rm asym}}\,(\Delta v)+d)}^2} \Big),
\end{equation} 
where $\Delta v$ is the Ly$\alpha$ velocity with respect to the systemic redshift, $A$ is the normalisation, $a_{\rm asym}$ the asymmetry parameter and where $d$ controls the line-width. By fitting to 1000 bootstrap-resamples of the data, we measure that Ly$\alpha$ flux peaks at $z=6.534\pm0.001$, which corresponds to a velocity offset of $+213^{+19}_{-20}$ km s$^{-1}$ to the systemic redshift traced by [CII]$_{158 \rm \mu m}$ ($z=6.5285$; \citealt{Matthee2019}, see Fig. $\ref{fig:spectra}$). The line-width FWHM is $374^{+21}_{-23}$ km s$^{-1}$ and the best-fitted asymmetry $a_{\rm asym}=0.34\pm0.03$. We measure an integrated Ly$\alpha$ luminosity of $(2.66\pm0.15)\times10^{43}$ erg s$^{-1}$, which corresponds to an EW$_{0}=38\pm5$ {\AA} when combined with the UV luminosity and slope (M$_{1500} =-22.37\pm0.05$, $\beta=-1.38^{+0.29}_{-0.27}$; \citealt{Matthee2019}). Our measurements from the MUSE, HST and ALMA data are summarised in Table $\ref{tab:properties}$. The Ly$\alpha$ luminosity is consistent with the slit-loss corrected flux (using the Ly$\alpha$ narrow-band) measured with X-SHOOTER \citep{Matthee2017SPEC}.

\begin{table}
\centering
\caption{Integrated measurements from the HST/WFC3, ALMA and MUSE data.} 
\begin{tabular}{lr}
\hline
Property & Value  \\  \hline
HST/WFC3 \& ALMA & \cite{Matthee2019} \\ 
R.A. & 22:18:56.36 \\
Dec. & +00:08:07.32 \\
M$_{1500}$ & -22.37$\pm0.05$ \\
$\beta$ & $-1.38^{+0.29}_{-0.27}$ \\
z$_{\rm sys, [CII]}$ & $6.5285^{+0.0007}_{-0.0004}$ \\
FWHM$_{\rm [CII]}$ & $200^{+48}_{-32}$ km s$^{-1}$ \\
SFR$_{\rm UV+IR}$ & $54^{+5}_{-2}$ M$_{\odot}$ yr$^{-1}$ \\
\hline
MUSE & This paper \\ 
$z_{\rm spec, Ly\alpha}$ & $6.534\pm0.001$ \\
$\Delta v_{\rm Ly\alpha}$ & $213^{+19}_{-20}$ km s$^{-1}$ \\
L$_{\rm Ly\alpha}$  & $(2.66\pm0.15)\times10^{43}$ erg s$^{-1}$ \\
EW$_{\rm 0, Ly\alpha}$ & $38\pm5$ {\AA} \\
FWHM$_{\rm Ly\alpha}$ & $374^{+21}_{-23}$ km s$^{-1}$ \\ 
$a_{\rm asym, Ly\alpha}$ & $0.34\pm0.03$ \\
$m_{\rm AB, 920-930 nm}$ & $25.16^{+0.40}_{-0.29}$ (S/N=3.8) \\
\hline
\end{tabular}
\label{tab:properties}
\end{table}

Compared to the galaxy population at $z\sim7$, VR7 has a high UV and Ly$\alpha$ luminosity ($\approx$5 L$^{\star}_{\rm UV}$ and $2.5\times$L$^{\star}_{\rm Ly\alpha}$, respectively). It is unclear whether the Ly$\alpha$ EW of VR7 is typical for its luminosity and redshift. \cite{Curtis-Lake2012} find that 50 \% of $>$L$^{\star}_{\rm UV}$ galaxies at $z=6.0-6.5$ have an EW$_0>25$ {\AA}, while \cite{Furusawa2016} find no strong Ly$\alpha$ emission in nine observed LBGs at a photometric redshift $z\sim7$. While the [CII]-UV luminosity ratio is representative of galaxies with similar luminosity (indicating a gas-phase metallicity $\approx0.2$ Z$_{\odot}$; \citealt{Matthee2019}), the upper limit on the IR-UV luminosity ratio (the latter constrained by $\lambda_0=160 \mu$m continuum observations) is extremely low, indicating faint IR luminosity and hence little obscured star formation yielding SFR$_{\rm UV+IR}=54^{+5}_{-2}$ M$_{\odot}$ yr$^{-1}$ \citep{Matthee2019}.

Despite its luminosity, there are no clear indications of AGN activity in VR7. The Ly$\alpha$ EW$_0$ is easily explained by a star-forming nature. No strong rest-frame high-ionisation UV lines as NV, CIV or HeII are detected \citep{Matthee2017SPEC}, the Ly$\alpha$ line is relatively narrow, significantly narrower than most (narrow-line) AGN \citep{Sobral2018}. Similar to other UV-luminous galaxies at $z\sim7$ \citep[e.g.][]{Bowler2017} VR7's rest-frame UV emission is resolved into two components with effective radii $0.84-1.12$ kpc and comparable luminosity. This is further evidence against dominant single nuclear emission. We can not distinguish whether the UV and [CII] components in VR7 correspond to two individual galaxies that are merging, or two star-forming complexes within the same galaxy. The separations in terms of projected distance ($\sim2$kpc) and velocity ($\sim200$ km s$^{-1}$) are relatively small. The resolved [CII] spectroscopy does not reveal ordered rotation on these scales \citep{Matthee2019}, indicating that the two components do not belong to the same dynamical system. Individually, the components have UV sizes that are comparable to other galaxies. Therefore, we conclude that VR7 is a relatively typical luminous star-forming galaxy at $z\sim6-7$ with the light being emitted roughly equally over two closely separated components.

\section{Environment around VR7} \label{app:environment}
We use the MUSE data to search for LAEs in the vicinity of VR7 in order to assess whether it resides in an extreme over-dense region. Specifically, we use CubEx to identify line-emitters in the continuum-subtracted data-cube observed at $\lambda_{\rm air} = 900-930$ nm. This wavelength range corresponds to $z=6.403-6.651$, or $\Delta$v$=\pm5000$ km s$^{-1}$ with respect to VR7's systemic redshift. The area of our data is 1.076 arcmin$^2$, meaning that our comoving survey volume is 606.6 cMpc$^3$. In order to identify reliable emission-lines, we require that at least 30 connected voxels have an individual S/N of 1.8 (thus integrated S/N$>10$). 

Our search results in 18 emission lines, with fluxes in the range $(2.7\pm0.3) - (87.5\pm0.3)\times10^{-18}$ erg s$^{-1}$ cm$^{-2}$. We use the full wavelength range observed with MUSE to identify the redshift of each emission line. We detect two emission lines for three objects in the investigated wavelength range (either H$\beta$+[OIII]$_{4959{\AA}}$ or [OIII]$_{4959{\AA}}$+[OIII]$_{5007{\AA}}$), meaning that 15 unique objects are found. A total of 10 objects are identified as [OIII] and/or H$\beta$ emitters at $z=0.82-0.92$, 3 objects are [OII] emitters at $z=1.40-1.48$, 1 object is part of a [SII] doublet at $z=0.34$ and 1 object is VR7 at $z=6.53$. Therefore, no neighbouring LAEs are detected around VR7 above a S/N$>10$.

The number of expected LAEs depends on the local (over)density and the luminosity function convolved with the completeness function of our data. We test the detection completeness of our data and methodology by injecting simulated LAEs in the data-cube and measuring the recovery fraction as a function of line-width and luminosity. We use {\sc imfit} \citep{Erwin2015} to simulate the spatial profile of LAEs with the typical Ly$\alpha$ surface brightness profile of LAEs at $z=5-6$ \citep{Wisotzki2018}, convolved with the PSF of our data. The Ly$\alpha$ flux is distributed over a half-gaussian line-profile (a gaussian with zero flux left-ward of the line-centre), which mimics the observed red asymmetric Ly$\alpha$ line profile of the majority of high-redshift LAEs. The line-profiles are smoothed with a gaussian line spread function with FWHM=85 km s$^{-1}$ at the observed wavelength \citep{Bacon2017}. We inject 20 simulated LAEs at random positions in the data cube and store the recovered fraction after applying our source detection methodology. This process is repeated 50 times to increase the statistics. We vary the line-widths from 100-500 km s$^{-1}$ in steps of 50 km s$^{-1}$, the total line-fluxes vary from $10^{-19}$ to $10^{-15}$ erg s$^{-1}$ cm$^{-2}$ and we vary the peak redshift from $z=6.40-6.60$. 

\begin{figure}
\includegraphics[width=8.6cm]{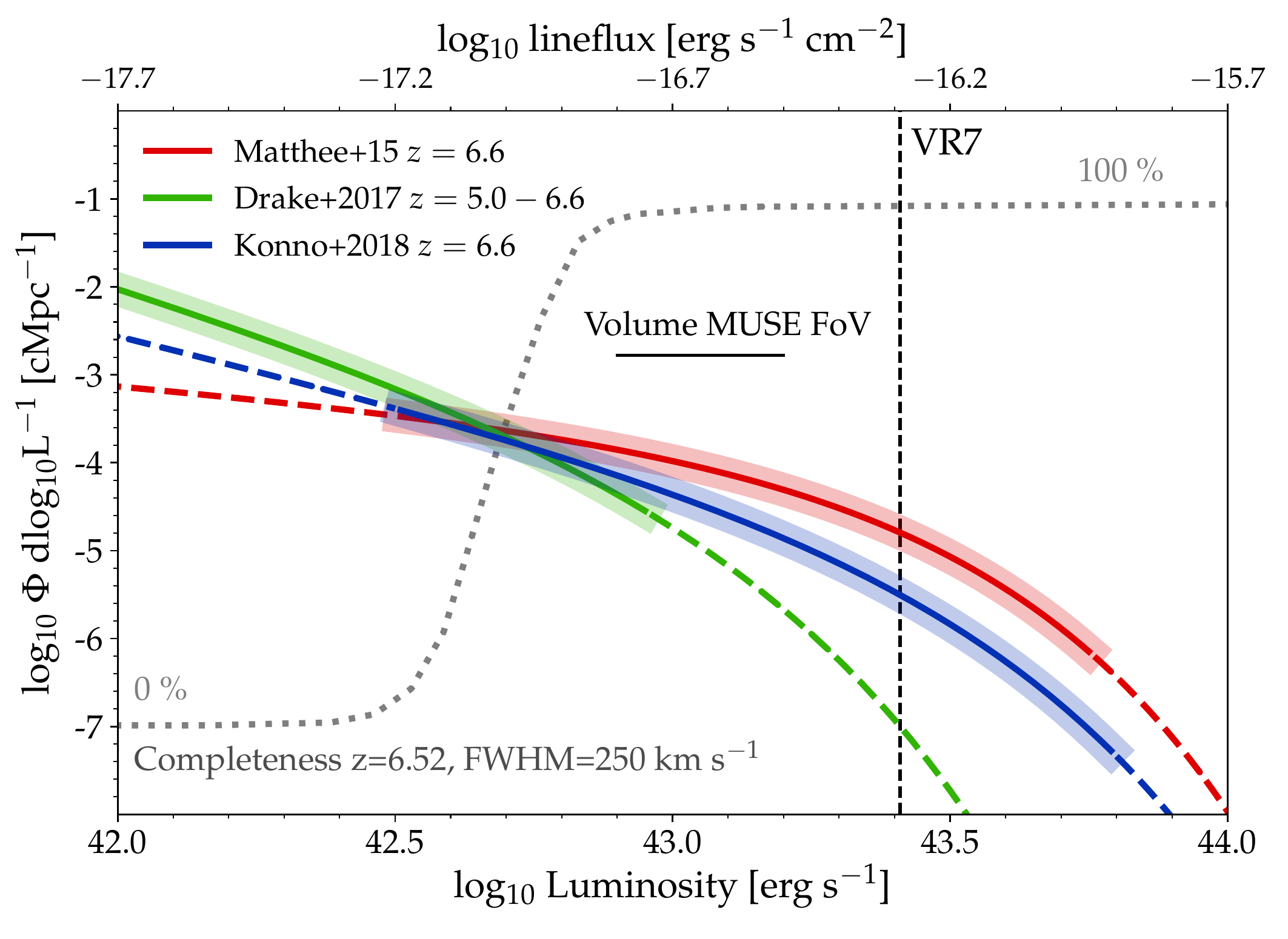}
    \caption{Ly$\alpha$ luminosity functions and completeness curve of our MUSE data. The blue and red LFs are from wide-field narrow-band surveys (\citealt{Matthee2015,Konno2018}) and the green LF is from a deep MUSE IFU survey (\citealt{Drake2017}). The line-style of the LFs changes to dashed in regimes where they are extrapolated, highlighting the complementarity of IFU and NB surveys. Shaded regions around the LFs highlight the uncertainties. The grey dotted line shows the completeness curve for simulated LAEs with line-width 250 km s$^{-1}$ at $z=6.52$. We also show the Ly$\alpha$ luminosity of VR7 and the volume probed at $z=6.403-6.651$ in our MUSE data.}
\label{fig:LF}
\end{figure}

\begin{figure*}
\includegraphics[width=17.8cm]{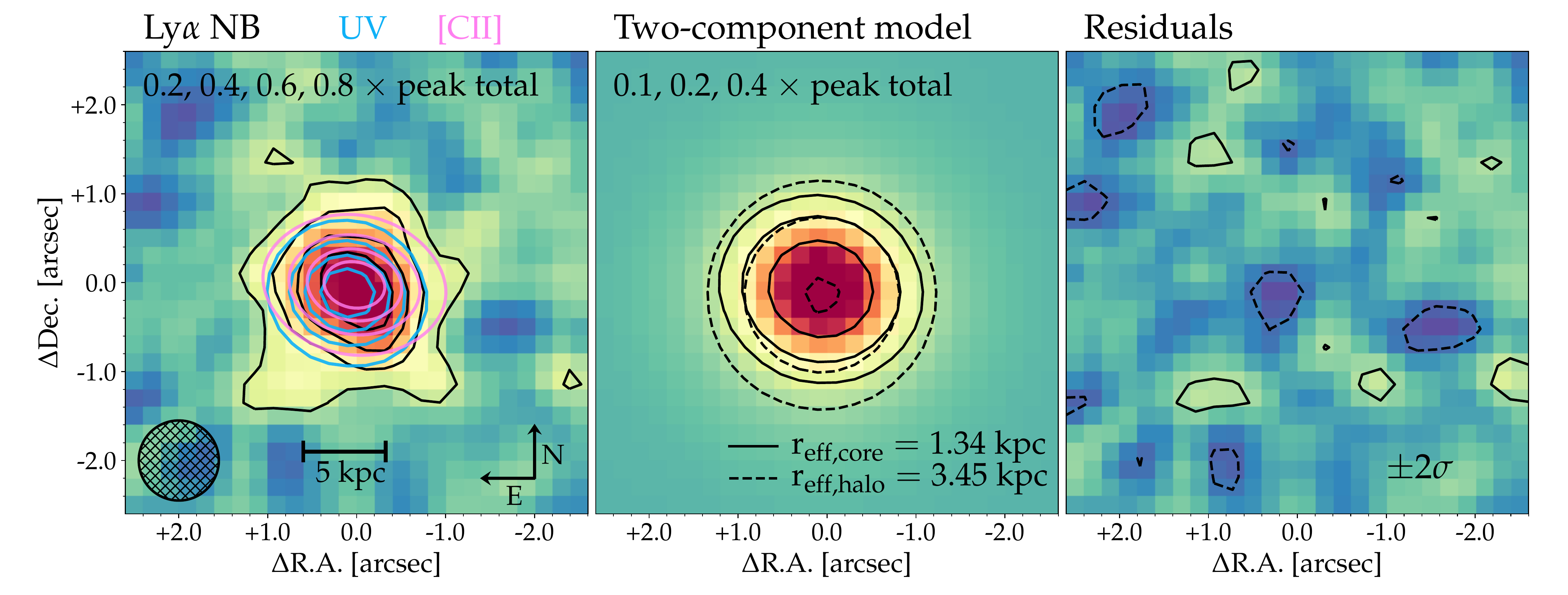}
    \caption{Continuum-subtracted Ly$\alpha$ image (left), best-fitted two-component model (middle) and the residuals (right). The Ly$\alpha$ image is constructed over $\lambda_{\rm obs, air}=9153.5-9170.0$ {\AA} in order to optimise the S/N, see Fig. $\ref{fig:spectra}$ and \S $\ref{sec:SB_Lya}$. The outer 0.2$\times$ peak flux contours in the left panel correspond to the 3$\sigma$ level. In the left panel, we also illustrate the rest-frame UV and [CII] morphology (from {\it HST}/WFC3 and ALMA; but convolved to the MUSE PSF as described in \S $\ref{sec:SB_HSTALMA}$, see \citealt{Matthee2019} for their images in higher resolution) in blue and pink, respectively.  The black hashed circle shows the MUSE PSF-FWHM. The two-component model (\S $\ref{sec:SB_2comp}$) is a combination of an exponential UV-like core component (illustrated with solid contours; based on {\it HST} data) and a spherically symmetric exponential halo component (dashed contours).}
\label{fig:halo_flux} 
\end{figure*} 

The detection completeness depends strongly on the assumed line-width: i.e. a 50 \% completeness (at $z=6.52$, but with little dependence on redshift) is achieved at L$_{\rm Lya}=2.8 (4.2)\times10^{42}$ erg s$^{-1}$ for a line-width of 100 (200) km s$^{-1}$, while 50 \% completeness is achieved only at  L$_{\rm Lya}=6.8\times10^{42}$ erg s$^{-1}$ for lines with 400 km s$^{-1}$. Previous observations of LAEs at $z\sim6.5$ show little dependence between line-width and luminosity \citep{Matthee2017SPEC}, with a typical line-width 250 km s$^{-1}$ (see also \citealt{Herenz2019}), which we use in our completeness estimate here.\footnote{We retrieve similar results when using a (weakly) luminosity-dependent line-width following \citep{Matthee2017SPEC}. The completeness function flattens in case line-widths follow a non-symmetric distribution (at fixed luminosity), but the shape and width of this distribution is currently not known and we therefore ignore scatter in line-widths at fixed luminosity. }

In Fig. $\ref{fig:LF}$ we show the Ly$\alpha$ luminosity function at $z=6.6$ derived from narrow-band surveys from \cite{Matthee2015} and \cite{Konno2018}, and the luminosity function at $z=5.0-6.6$ from deep MUSE observations \citep{Drake2017}. Although the luminosity functions show some differences at the bright and faint end\footnote{Note that the overlap in dynamic range between narrow-band surveys (which fixed the faint-end slope) and MUSE observations are limited.}, their intersection point coincides with the luminosity at which our MUSE data is $\approx50$\% complete. As a consequence the predicted number of observed LAEs in the VR7 cube is not very sensitive to the differences in the luminosity functions, particularly since the VR7 cube probes a small cosmic volume. By integrating the luminosity functions convolved with the completeness curve, we find that 0.06-0.10 LAEs are expected to be identified between $z=6.403-6.651$ in the VR7 cube if it has a normal density. Since we do not identify a LAE besides VR7, we can thus exclude that the environment of VR7 is more than 10 times over-dense. Therefore, our MUSE data is not deep (and/or wide) enough to be used to identify moderate over-densities.

\section{Sizes and extended emission} \label{sec:SB} 
In this section, we investigate the extent of VR7's Ly$\alpha$ emission in the MUSE data and compare this to the rest-frame UV and [CII] sizes. 

\subsection{Ly$\alpha$} \label{sec:SB_Lya}
We create a continuum-subtracted Ly$\alpha$ narrow-band image from the MUSE data-cube using a {\it HST}-based continuum model, convolved with the MUSE PSF with {\sc Imfit}, see Appendix $\ref{app:UVcont}$ for details. The continuum-subtracted Ly$\alpha$ image is shown in the left panel of Fig. $\ref{fig:halo_flux}$. Due to the high {\it observed} Ly$\alpha$ EW (EW$_{\rm obs} = (1+z)$ EW$_0 =286$ {\AA}), the continuum subtraction does not impact the Ly$\alpha$ morphology significantly (the continuum contributes $\approx8$ \% of the flux in the narrow-band over $\lambda=9153.5-9170$ {\AA}). We note that our continuum model is consistent with the continuum image from the MUSE data at 920-930 nm (Appendix $\ref{app:UVcont}$).

Ly$\alpha$ emission is well resolved in our MUSE data. We use {\sc Imfit} to fit an exponential profile to VR7's Ly$\alpha$ image. In short, {\sc Imfit} simulates PSF-convolved models (using the reference star image) and finds the best-fitting solution using the Levenberg-Marquardt algorithm. Uncertainties are estimated by performing the fitting procedure on 1000 bootstrap-resamples of the data using the propagated pixel-uncertainties and deriving 68.4 \% confidence intervals. We do not impose spherical symmetry in these single component fits. We measure r$_{\rm eff, Ly\alpha}=2.05\pm0.16$ kpc with I$_{\rm eff, Ly\alpha}=(3.39\pm0.45)\times10^{-17}$ erg s$^{-1}$ cm$^{-2}$ arcsec$^{-2}$, the surface brightness at the effective radius. The best-fit ellipticity is $0.15^{+0.14}_{-0.06}$ with PA$=189^{+18}_{-33}$ degrees. If we allow the S\'ersic index to vary between $n=0.01$ and $n=10$ we find a best-fit index $n=0.55\pm0.42$ with a slightly larger scale radius, r$_{\rm eff, Ly\alpha}=2.17\pm0.20$ kpc and similar ellipticity and position angle as the exponential model. Our results are listed in Table $\ref{tab:properties_size}$.

\subsection{UV and [CII]$_{158\mu{\rm m}}$} \label{sec:SB_HSTALMA}
For a proper comparison to the rest-frame UV ({\it HST}/WFC3) and [CII] (ALMA) data, we use single exponential profiles to describe their morphologies using modelled images with the same PSF as the MUSE data. In their spatial resolution of $0.25''$ and $0.5''$, respectively, the  morphology in both UV and [CII] is well described by a combination of two exponential components \citep{Matthee2019} that are oriented in the east-west direction. However, most of this structure is not seen in the MUSE resolution due to the larger PSF (see the contours in the left panel of Fig. $\ref{fig:halo_flux}$). Fitting the convolved UV and [CII] images with a single exponential profile  as in \S $\ref{sec:SB_Lya}$, we measure (de-convolved) effective radii  r$_{\rm eff, UV}=1.34\pm0.06$ kpc and r$_{\rm eff, [CII]}=2.14^{+0.24}_{-0.22}$ kpc (where the errors include propagating the uncertainties in the fits to the morphologies at higher resolution). The UV scale length is significantly smaller than the Ly$\alpha$ scale length (by a factor $1.5\pm0.1$), but the [CII] scale length is similar to Ly$\alpha$, see Table $\ref{tab:properties_size}$. Note that we do not force spherical symmetry on our UV and [CII] models.

\begin{figure}
\includegraphics[width=8.6cm]{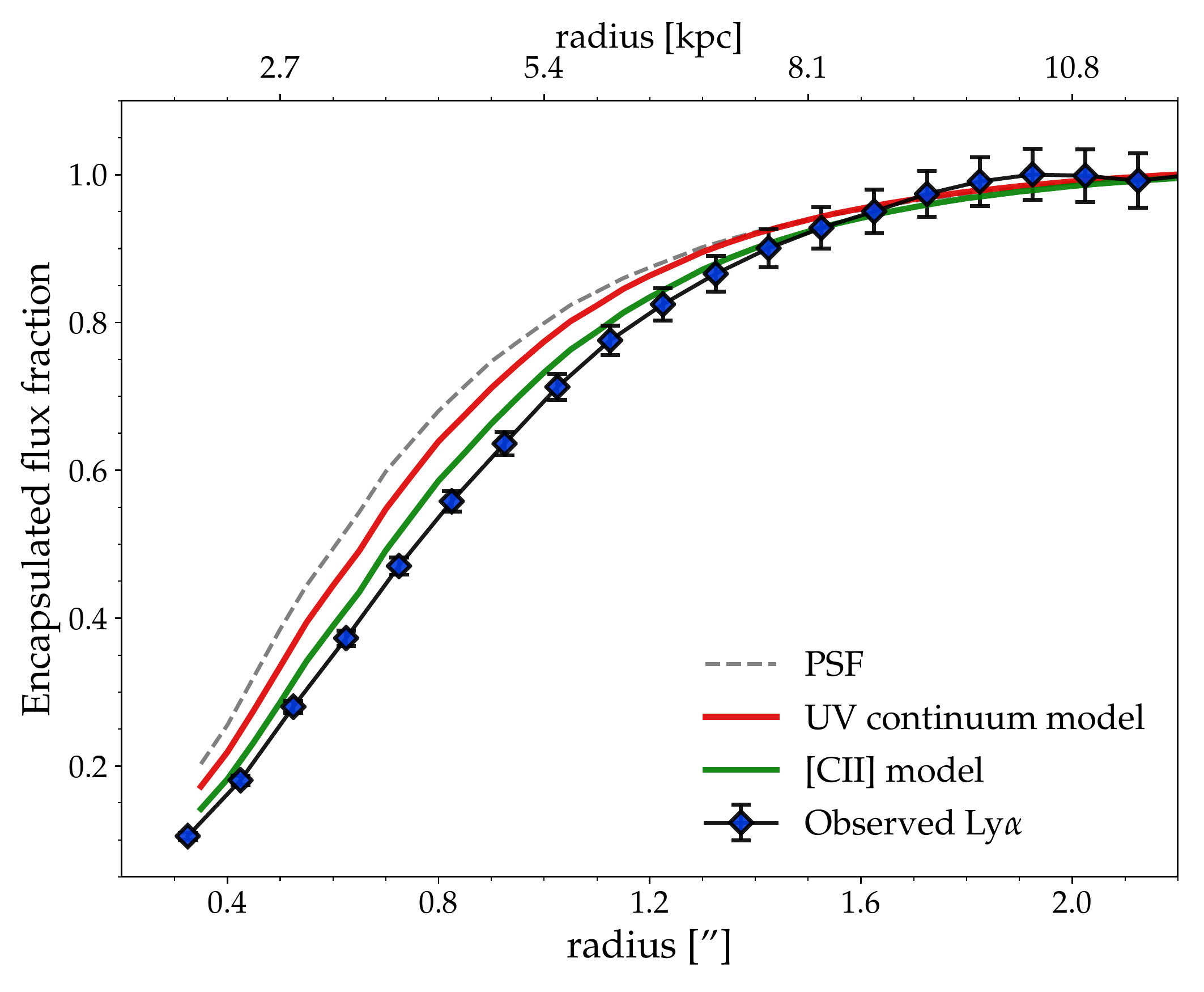}
    \caption{Curve-of-growth of VR7's Ly$\alpha$ emission (blue data-points), the {\it HST} based UV continuum model and ALMA based [CII] model convolved through the MUSE PSF (red and green lines, respectively) and the PSF in the MUSE data-set (grey dashed line). The Ly$\alpha$ emission, the [CII] emission and the UV continuum emission are extended, with Ly$\alpha$ having the largest scale length. }
\label{fig:growth}
\end{figure}

\subsection{Two-component Ly$\alpha$ fit} \label{sec:SB_2comp}
The relatively large Ly$\alpha$ scale length compared to the UV continuum indicates the presence of a Ly$\alpha$ halo. Following the analyses by e.g. \cite{Steidel2011} and \cite{Wisotzki2015}, we describe the Ly$\alpha$ surface brightness profile as the combination of a `UV-like' component (the core-component; the MUSE-PSF-convolved {\it HST} model described above) that dominates in the centre and an exponential component that dominates at large radii (the halo-component). The centroids of the single-component Ly$\alpha$ models (\S $\ref{sec:SB_Lya}$) and the core-component are separated by only $0.04\pm0.01''$ ($0.24\pm0.09$ kpc). Combined with the astrometric uncertainty of $0.02''$ this means there are no significant spatial offsets. Therefore, we assume the core and halo components to be co-spatial in our modelling below.

We use {\sc imfit} to simulate a range of model images in a grid of halo flux, core flux and halo effective radius and then we calculate the $\chi^2$ of each model compared to the continuum-subtracted Ly$\alpha$ image. For each model we also calculate the likelihood as $\mathcal{L} \propto {\rm exp}(-\chi^2/2)$. Each model consists of an exponential core component (with effective radius fixed to the UV size; r$_{\rm eff, core}=1.34$ kpc, ellipticity 0.53 and position angle PA=74 degree) and a spherically symmetric halo component with effective radius r$_{\rm eff, halo}>1.34$ kpc (Table $\ref{tab:properties_size}$). We find a best-fit halo effective radius r$_{\rm eff, halo} = 3.45^{+1.08}_{-0.87}$ kpc and a halo flux fraction of $54^{+11}_{-10}$ \%. Here, the quoted values represent the median and the difference to the 16th and  84th percentile of the marginalised posterior distribution. Our results indicate that the majority of Ly$\alpha$ has a significantly different morphology from the UV. The best fit and its residuals are illustrated in Fig. $\ref{fig:halo_flux}$.

\begin{figure}
\includegraphics[width=8.6cm]{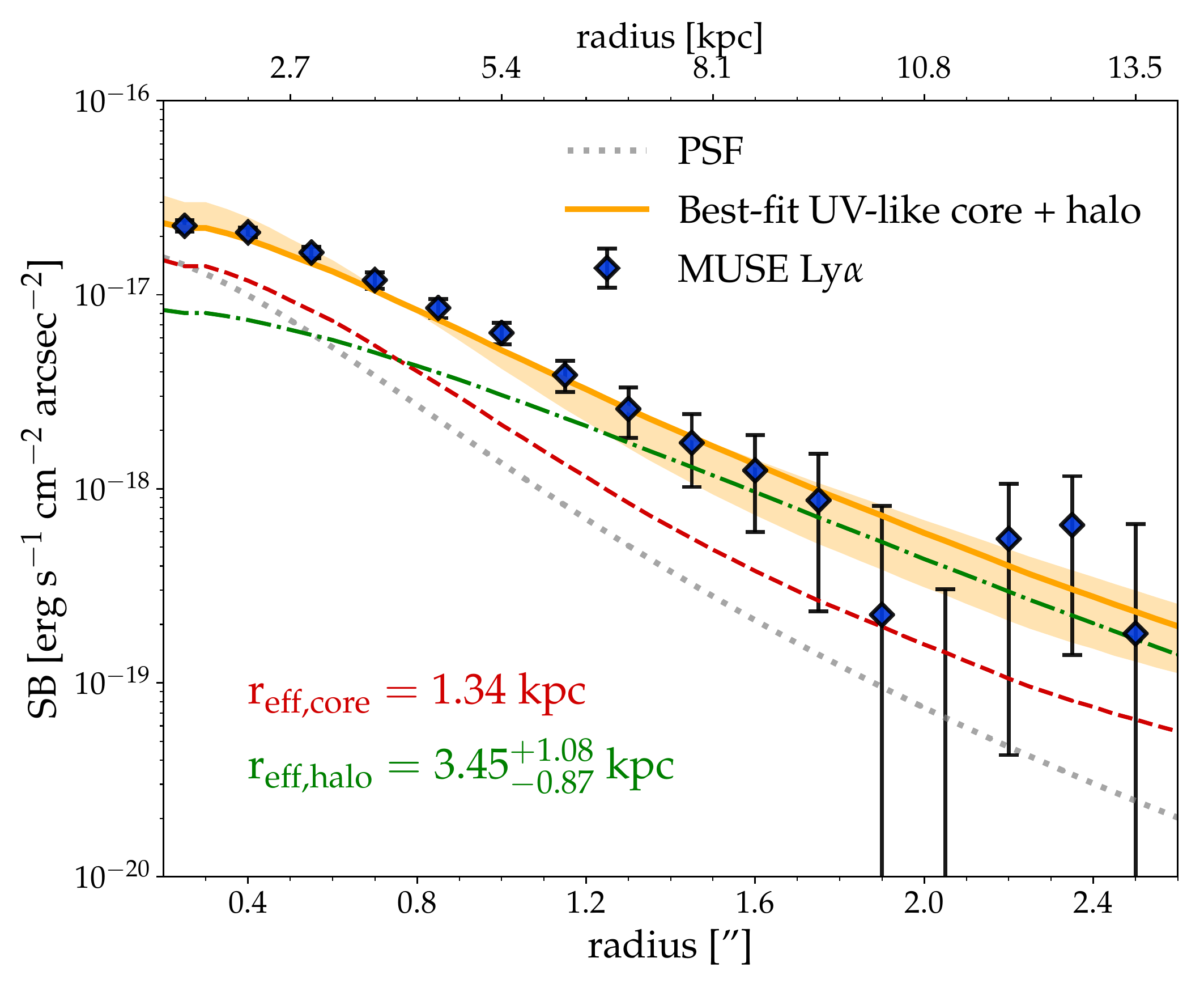}
    \caption{One-dimensional Ly$\alpha$ surface brightness profile of VR7 (blue points) and the best-fitted two component model and its uncertainties (orange line and shaded region). The dashed and dashed-dotted lines show the two components contributing to the Ly$\alpha$ SB profile, which consist of a core-like profile (red; with the same scale length as the UV continuum) and a halo-like profile (green; with larger scale length by definition), see \S $\ref{sec:SB_2comp}$. The grey dotted line shows the PSF. Note that while the shown SB profiles are the observed profiles (after PSF-convolution), our listed scale lengths are deconvolved values. With the current data, halo-like Ly$\alpha$ flux dominates at radii $>0.8''$.  }
\label{fig:SB_DECOMP}
\end{figure}

\subsubsection{1D SB profile} \label{sec:SB_1D}
To facilitate comparison with literature results, we now focus on the (spherically averaged) 1D growth curve and surface brightness (SB) profile. We extract the 1D SB profile by measuring the summed flux in increasingly large concentric annuli, divided by the area of each annulus. Annuli are centred at the location of peak emission. The maximum annulus has a radius of 2$''$. The errors on the 1D SB profile are estimated by extracting the SB profile in each of the 62 empty sky regions (see \S $\ref{sec:quality}$) and computing the standard deviation of the SBs in each annulus. 

The observed growth curve of VR7's Ly$\alpha$ emission is shown in Fig. $\ref{fig:growth}$, where we also show the curves of the reference star used for the measurement of the PSF and of the (convolved) UV continuum and [CII] models. At the resolution of our MUSE observations VR7 is spatially resolved in UV, [CII] and Ly$\alpha$ emission, with increasing scale lengths, respectively. This is similar to the result by \cite{Fujimoto2019}, who find a larger [CII] scale length than the UV scale length using stacks of galaxies at $z\sim6-7$. In Fig. $\ref{fig:SB_DECOMP}$ we show how the 1D SB profile of VR7's Ly$\alpha$ emission is decomposed into the core and halo-components. These 1D SB profiles are measured on the best-fitted two-dimensional models described above. Halo flux overtakes that of the core Ly$\alpha$ flux at radii $>0.8''$.

\begin{table}
\centering
\caption{Morphological measurements. The rest-frame UV measurement is performed on {\it HST} data that has been PSF-matched to the MUSE data.} 
\begin{tabular}{lr}
\hline
Property & Value  \\  \hline
MUSE Ly$\alpha$ & \\ 
Single component  & \\
r$_{\rm eff, Ly\alpha, exponential}$  & $2.05\pm0.16$ kpc\\
ellipticity$_{\rm eff, Ly\alpha, exponential}$ & $0.15^{+0.14}_{-0.06}$ \\
PA$_{\rm eff, Ly\alpha, exponential}$ & $189^{+18}_{-33}$ \\
r$_{\rm eff, Ly\alpha, Sersic}$  & $2.17\pm0.20$ kpc\\
 ellipticity$_{\rm eff, Ly\alpha, Sersic}$ & $0.17^{+0.11}_{-0.05}$ \\
 PA$_{\rm eff, Ly\alpha, Sersic}$ & $189^{+17}_{-25}$ \\
$n_{\rm Sersic}$ & $0.55\pm0.42$\\ 
UV-like core + spherical Ly$\alpha$ halo & \\
r$_{\rm eff, Ly\alpha, halo}$  & $3.45^{+1.08}_{-0.87}$ kpc\\ \hline
UV &  \\
r$_{\rm eff, UV, core}$  & $1.34\pm0.06$ kpc\\ 
ellipticity$_{\rm UV}$ & $0.53\pm 0.05$ \\
PA$_{\rm UV}$ & $74\pm4$ \\ \hline
[CII] &  \\ 
r$_{\rm eff, [CII]}$  & $2.14^{+0.24}_{-0.22}$ kpc\\ 
ellipticity$_{\rm [CII]}$ & $0.76\pm 0.12$ \\
PA$_{\rm [CII]}$ & $80\pm6$ \\ \hline
\end{tabular}
\label{tab:properties_size}
\end{table}

\section{Resolved Ly$\alpha$ properties} \label{sec:linevariations}
At the spatial resolution of the ALMA and {\it HST}/WFC3 data, VR7 is observed to consist of multiple components, while at the MUSE resolution the Ly$\alpha$ emission appears to be described by a single component in a standard collapsed pseudo-NB image. By exploiting the full 3D information of the MUSE data, together with higher spatial resolution ALMA and {\it HST} data, we explore whether multiple components can also be related to the Ly$\alpha$ emission, particularly in the line of sight.

\begin{figure*}
\begin{tabular}{cc}
\includegraphics[width=8.0cm]{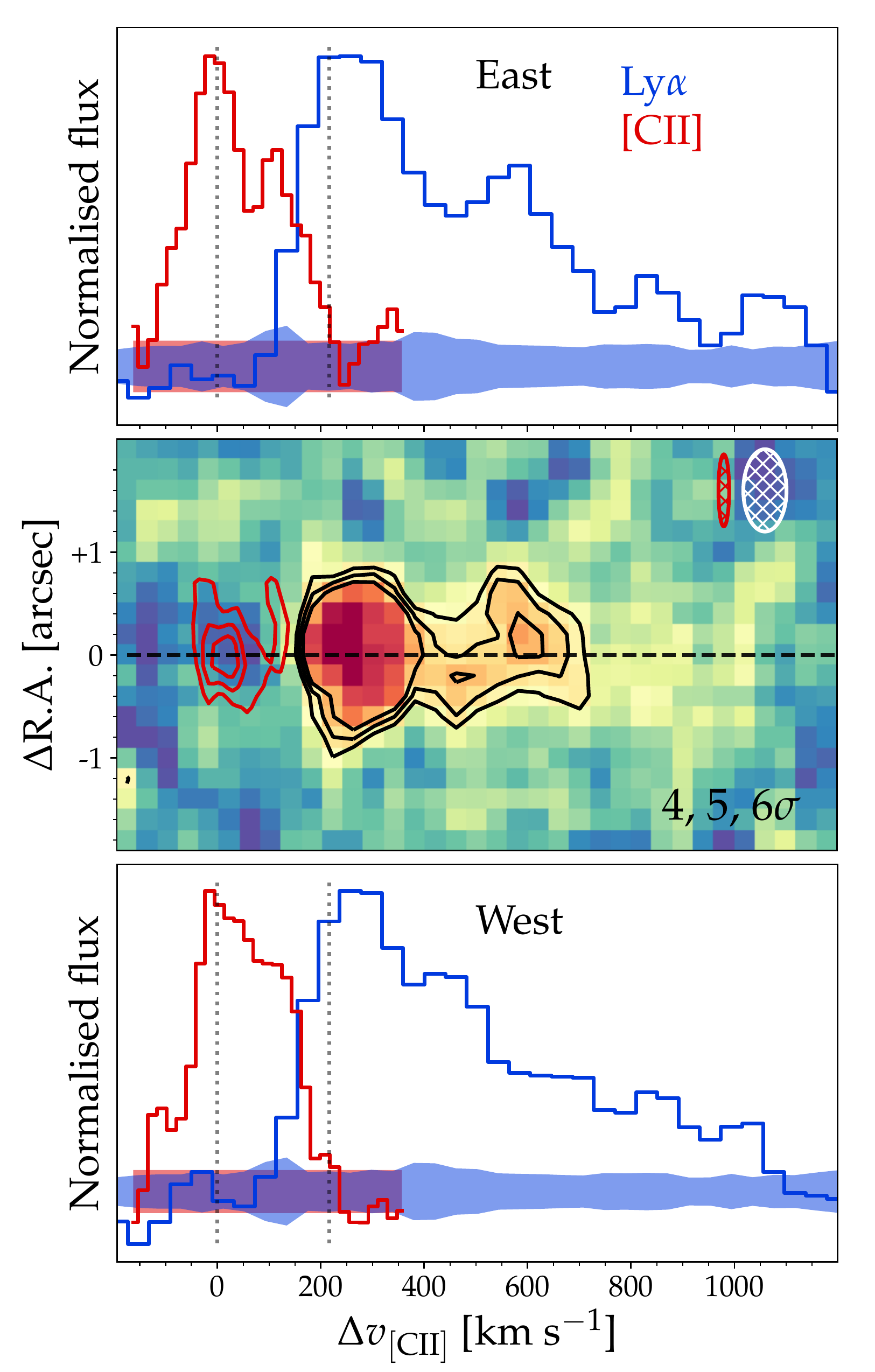} & 
\includegraphics[width=8.0cm]{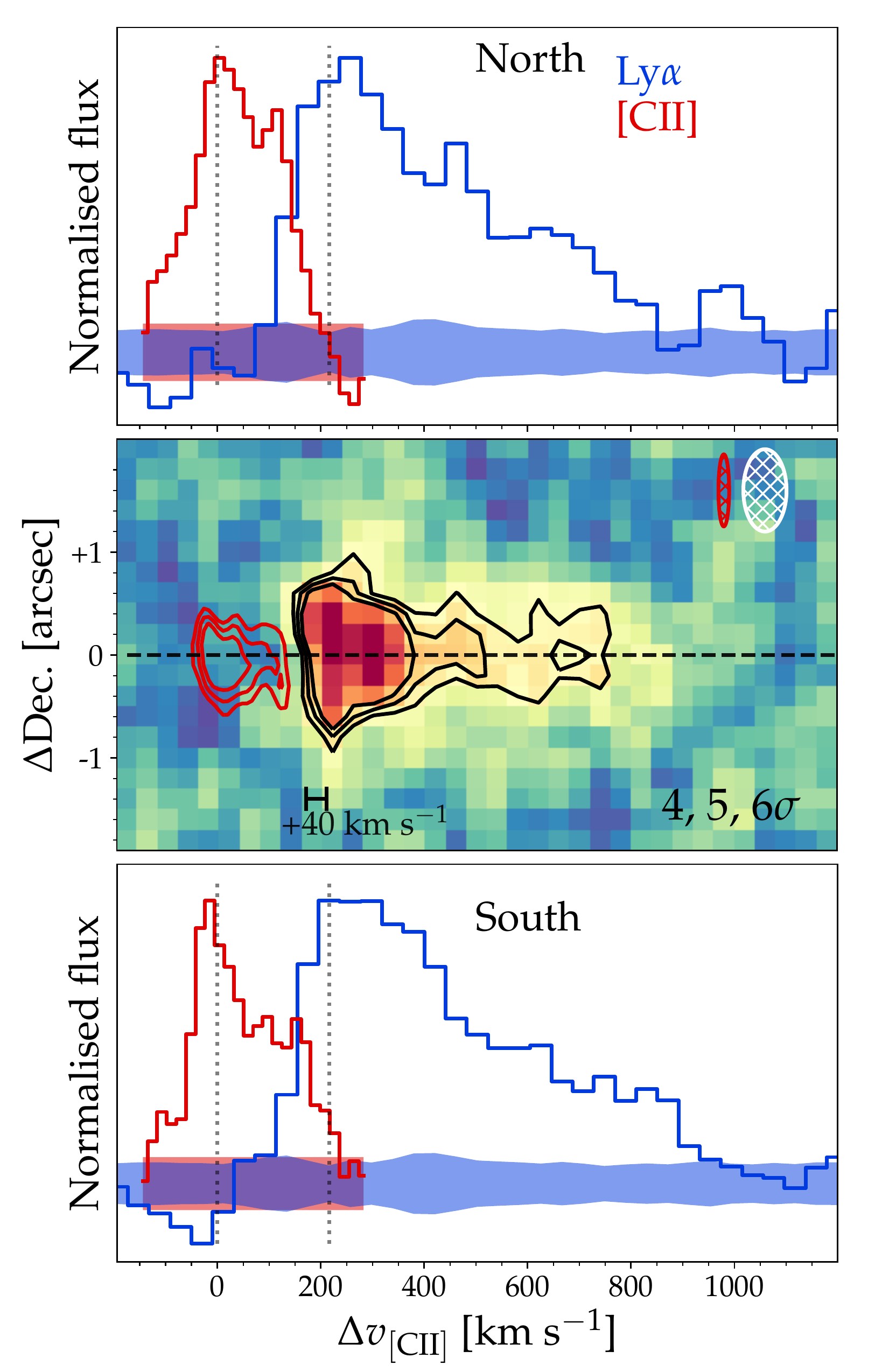} \\
\end{tabular}
    \caption{Spectral variations of [CII] and Ly$\alpha$ for VR7, depending on R.A. (left) and Dec. (right). Top and bottom rows show 1D extractions in Ly$\alpha$ (blue, extracted from the centre to $\pm1''$) and [CII] (red, from ALMA; \citealt{Matthee2019}). The middle row shows the Ly$\alpha$ pseudo-slit, where black (red) contours mark the 4, 5, 6$\sigma$ levels of the MUSE (ALMA) data. The white (red) ellipses show the PSF and LSF of the MUSE (ALMA) data. A second Ly$\alpha$ component is clearly visible towards the east, while the western component is broader. The peak of the Ly$\alpha$ line in the south/west is tentatively shifted by $\approx40$ km s$^{-1}$ compared to the north/east.}
\label{fig:PV}
\end{figure*}

\begin{figure}
\includegraphics[width=8.4cm]{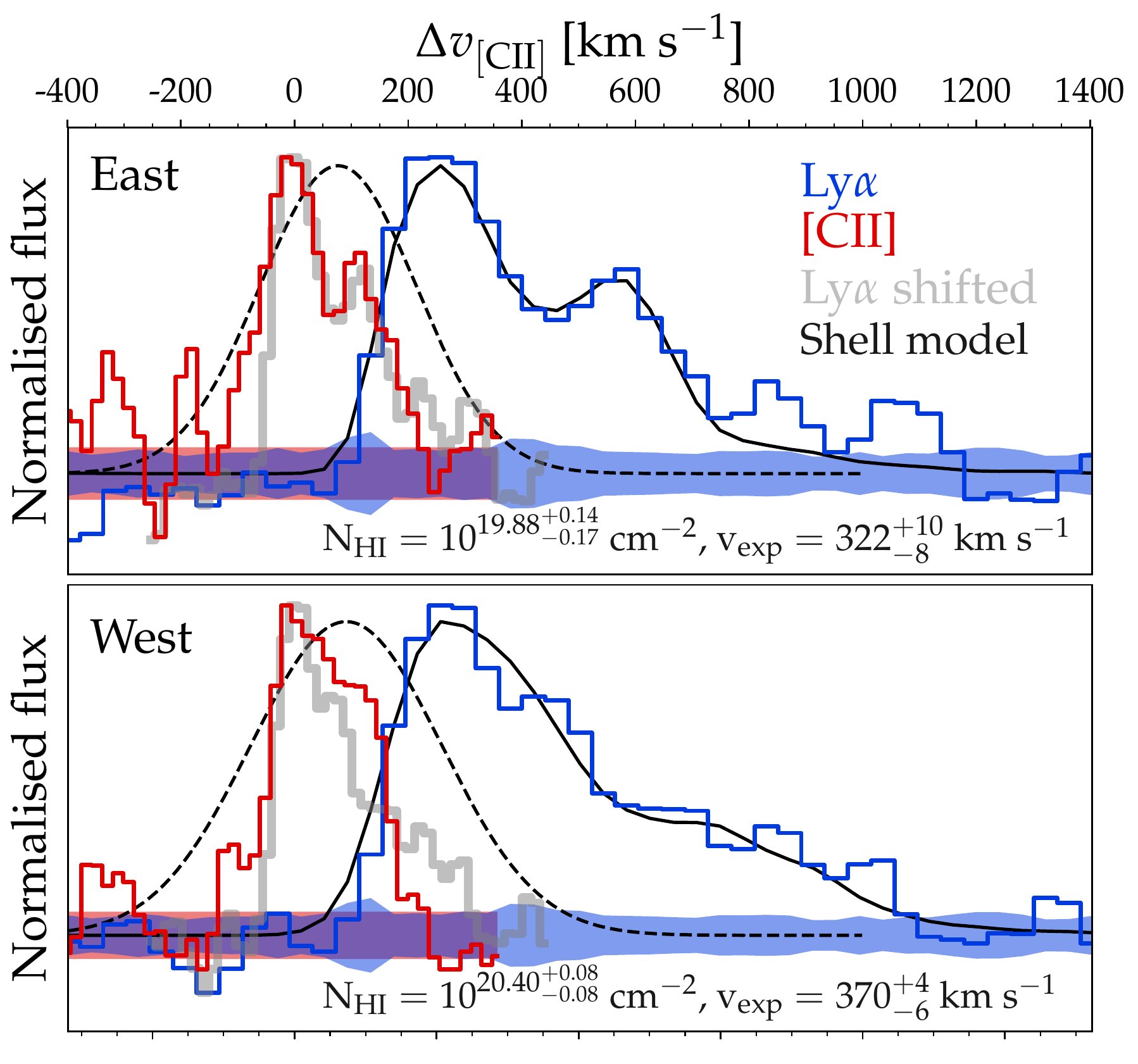} \vspace{-0.6cm}\\ 
\includegraphics[width=8.4cm]{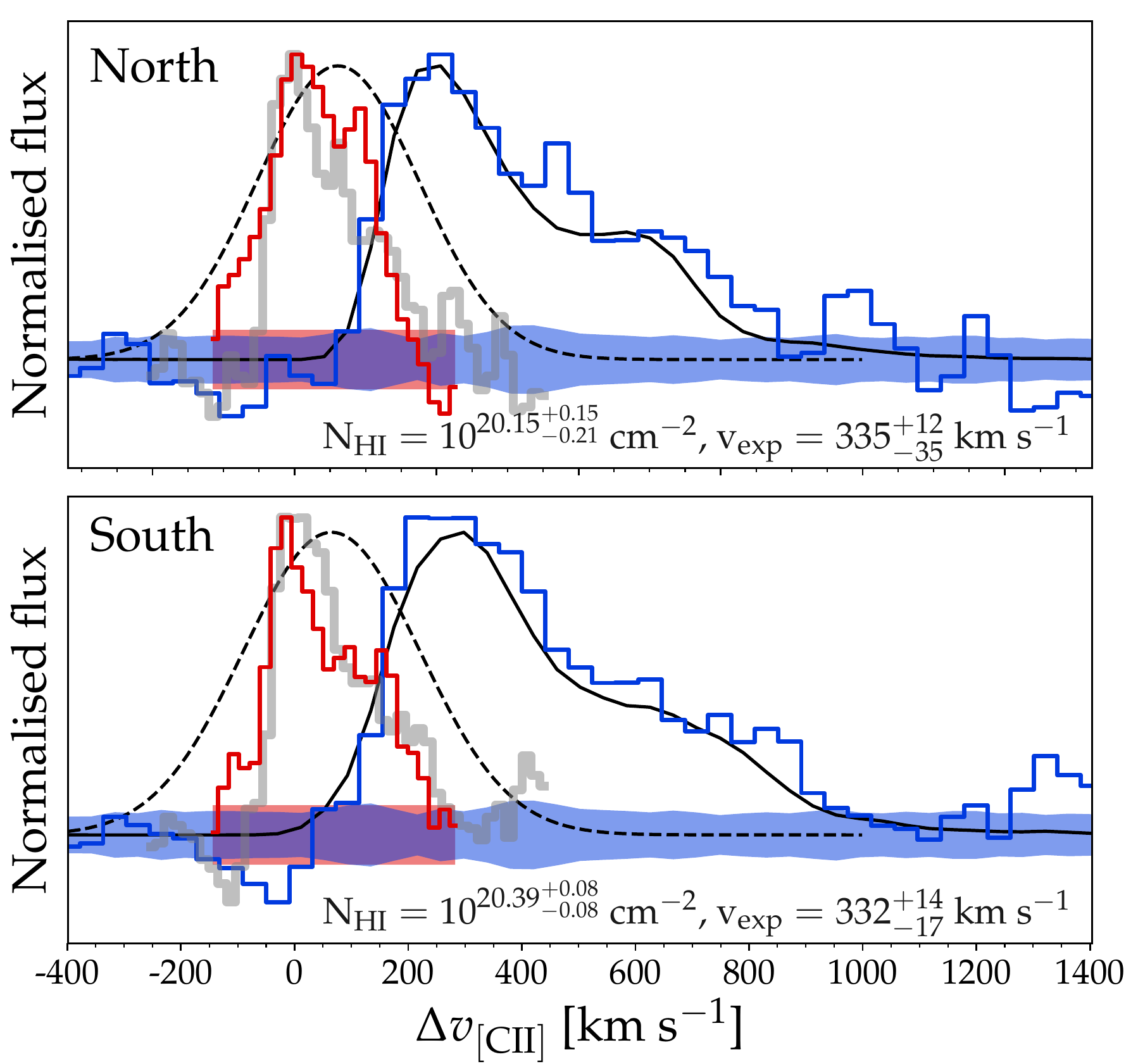} \\
    \caption{1D extractions of the spectral variations of [CII] (red) and Ly$\alpha$ (blue) for VR7, depending on position (as Fig. $\ref{fig:PV}$). The solid black lines show the best-fitted Ly$\alpha$ shell models and the dashed black lines show the intrinsic spectrum in those models. The intrinsic spectrum in the shell model fits is somewhat broader and redshifted compared to [CII]. We show in light-grey an artificial Ly$\alpha$ profile shifted as $v_{\rm new} = (v_{\rm obs}-260$ km s$^{-1})/2.6$ km s$^{-1}$ (see \S $\ref{sec:nature}$). It is remarkable that it resembles the [CII] profile in the east so well, but this could be a coincidence.}
\label{fig:shellmodel}
\end{figure}

\subsection{Ly$\alpha$ line-profile variations} \label{sec:profile_variations}
Here we investigate spatial variations in the Ly$\alpha$ line-profile and how these variations correlate with spatial variations in the [CII] emission line profile (Fig. 15 in \citealt{Matthee2019}). We explore spatial variations in the Ly$\alpha$ line profile by using position-velocity diagrams (PV diagrams; i.e. pseudo-slits) extracted over different regions of the galaxy. The benefit of PV diagrams is that they increase the S/N (by averaging over multiple pixels) without parametrising the data.

Fig. $\ref{fig:PV}$ shows PV diagrams in two halves of VR7. The extractions are centred on the peak Ly$\alpha$ emission and averaged over a 1.0$''$ slice in the orthogonal direction. As a reference velocity, we use the flux-weighted [CII] redshift, $z=6.5285$ \citep{Matthee2019}, which we refer to as the `systemic' redshift below. The left panel in Fig. $\ref{fig:PV}$ shows variations in the east-west direction, while the right panel shows variations in the north-south direction. The central row shows the PV diagrams, while the top and bottom rows show the 1D spectra by summing the PV diagrams $\pm1''$ from the centre.  

Two distinct spatial variations in the Ly$\alpha$ line profile can be identified in Fig. $\ref{fig:PV}$, which interestingly can also be identified in the [CII] spectra from ALMA (here imaged with spatial resolution with PSF-FWHM $\approx0.7''$; \citealt{Matthee2019}). In general, Ly$\alpha$ is redshifted by $\approx220$ km s$^{-1}$ compared to the [CII] line. In the east however, we observe a second bump in the Ly$\alpha$ emission line at $\approx+600$ km s$^{-1}$ with respect to the systemic redshift, while the Ly$\alpha$ line in the west is broader. The Ly$\alpha$ line FWHM in the east is $230^{+80}_{-40}$ km s$^{-1}$, while it is $360^{+20}_{-40}$ km s$^{-1}$ in the west. The second bump in the east is also seen in [CII] emission at a redshift of $\approx+130$ km s$^{-1}$ compared to the systemic. In [CII] emission, the second bump has a smaller velocity difference to the main component than the second Ly$\alpha$ bump has compared to the Ly$\alpha$ peak. There is a small tentative gradient in the peak-velocity of both the [CII] and Ly$\alpha$ lines. The peak shifts by $\approx+40$ km s$^{-1}$ from east to west and from north to south (see Fig. 14 in \citealt{Matthee2019} for a [CII] moment map that shows this more clearly). 

\begin{figure*}
\includegraphics[width=15.7cm]{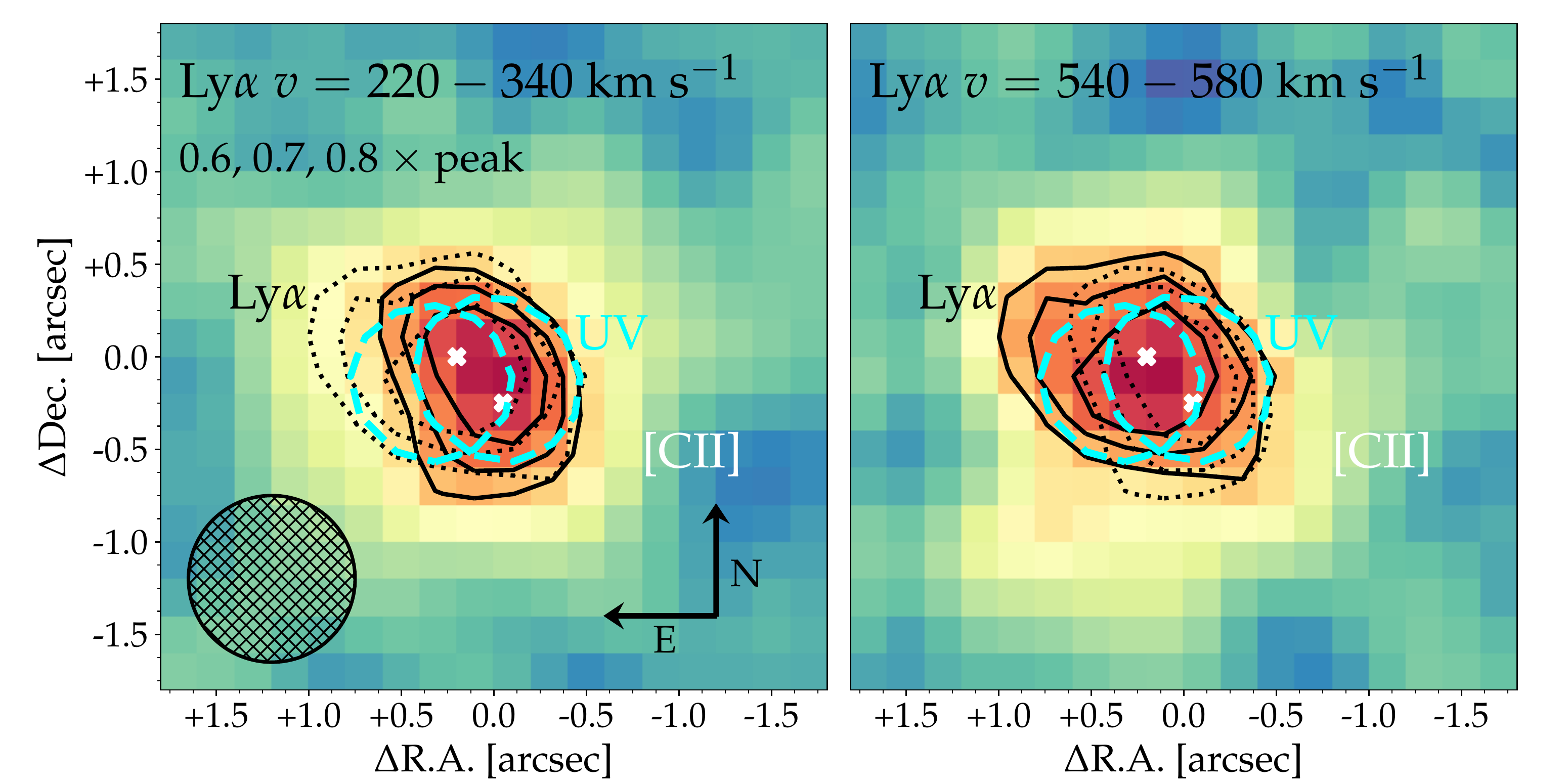}
    \caption{Ly$\alpha$ narrow-band images centred on the main peak (left) and on the ``bump" (right) identified in the PV diagrams (Fig. $\ref{fig:PV}$). Solid black contours mark the 0.6, 0.7 and 0.8 $\times$ the peak flux in each collapsed image, while the dotted black contours illustrate the contour levels from the other component, respectively. The black hashed circle shows the PSF of the MUSE data. The dashed cyan contours show the 0.6$\times$ peak UV flux in MUSE-PSF convolved model images of the individual components. White crosses mark the peak positions of the [CII] components.}
\label{fig:Lya_2comp}
\end{figure*}

\subsection{Ly$\alpha$ shell models}
In order to interpret the spatial variations in the Ly$\alpha$ profile using Ly$\alpha$ information alone, we perform `shell-model' fitting on continuum-subtracted Ly$\alpha$ spectra in different regions (the four directions in the PV-diagram; Fig. $\ref{fig:PV}$) of VR7 using the pipeline described in \citet{Gronke2015}. The shell-model is a popular model in order to extract physical information of Ly$\alpha$ spectra. It consist out of a Ly$\alpha$ and continuum emitting source which is surrounded by a shell of neutral hydrogen, and dust \citep{Ahn2002,Verhamme2006}. The model features a minimum of six free parameters: the width and equivalent width of the intrinsic Ly$\alpha$ line ($\sigma_{\mathrm{i}}$, $EW_{\mathrm{i}}$), the dust and hydrogen content of the source (which we characterize by the all-absorbing dust optical depth $\tau_{\mathrm{a}}$, and the neutral hydrogen column density of the shell $N_{\rm HI}$), the `effective temperature' of the shell $T$, and the inflow / outflow velocity of the shell $v_{\mathrm{exp}}$.
The Bayesian fitting pipeline used features the possibility to add additional parameters. We also leave the systemic redshift as an additional free parameter on which we impose the gaussian prior $(\mu,\,\sigma)=(6.5285,\,0.001)$ based on [CII]. We note that leaving $z_{\mathrm{sys}}$ free to vary is important as even a small shift can cause a sharp drop in likelihood. We also note that the LSF is incorporated in the fitting pipeline by smoothing the synthetic spectrum before computing the likelihood.

We show the best fitted models in Fig. $\ref{fig:shellmodel}$ (as black, solid lines in each panel). The shell model can describe the data in various positions of the galaxy very well. The systemic redshift obtained from the fit is consistent with the prior [CII] spectrum, but it is at slightly higher redshift than the [CII] peak (by +70 km s$^{-1}$). This could be due to the fact that VR7 is a two-component galaxy, see \S $\ref{sec:2comps}$. The intrinsic spectrum of the shell model fits is in all cases somewhat broader than the [CII] spectrum. By analysing the shell-model spectra, we found that the secondary red peak (particularly prominent in the east) consists mainly of so-called `backscattered' photons, which are photons with a last scattering angle $\cos\theta\sim -1$. These photons experience a boost of $\sim 2$ times the shell outflow velocity, thus leaving a characteristic `hump' in the emergent Ly$\alpha$ spectrum \citep{Ahn2002}. The best-fit shell model parameters vary from $N_{\rm HI} \sim 10^{20}$cm$^{-2}$ in the east to $N_{\rm HI}\sim 10^{20.4}$ cm$^{-2}$ in the west, and the outflow velocities from $320$ km s$^{-1}$ to $360$ km s$^{-1}$, respectively. These parameters are larger than those found in the luminous LAE CR7 \citep{Dijkstra2016}. We note that the physical nature of the shell-model parameters is still debated \citep[e.g.][]{Orlitova2018}. One caution is that Ly$\alpha$ photons trace preferably the low-$N_{\rm HI}$ medium, and thus, the spectral information does not necessarily correspond to the line-of-sight physical conditions \citep{Eide2018,KakiichiGronke2019}. We interpret these results in \S $\ref{sec:nature}$.

\subsection{Two Ly$\alpha$ emitting components?} \label{sec:2comps}
One explanation of the resemblance between the spatial variations in the [CII] and Ly$\alpha$ line profiles could be that the Ly$\alpha$ emission in VR7 also consists of components in the line of sight with slight velocity differences, each emitting Ly$\alpha$. 
 
In order to optimally isolate the two components identified in the PV diagram spatially, we collapse the MUSE data-cube over customized wavelength layers. Specifically, we collapse the layers from $v=220-340$ km s$^{-1}$ (where $v=0$ is the systemic redshift $z=6.5285$) to isolate the main peak and $v=540-580$ km s$^{-1}$ to isolate the redshifted bump, see Fig. $\ref{fig:Lya_2comp}$. While the large PSF significantly challenges the analysis, it is clear that the emission in the redshifted bump extends more towards the east, while the main component extends towards the south-west. Fitting the light-distribution with a single exponential similar to \S $\ref{sec:SB_Lya}$, we find that the peak position of the redshifted bump is shifted by $0.14\pm0.04''$ towards the east compared to the main component. Note however that the red asymmetric wing of the main component likely contributes to the image of the bump emission, indicating the real separation can be higher. The redder Ly$\alpha$ component is tentatively more extended (with an effective radius of $2.71\pm0.35$ kpc compared to $2.14\pm0.26$ kpc). 

For comparison, in Fig. $\ref{fig:Lya_2comp}$, we also show the (MUSE PSF-convolved) contours of the individual components identified in the rest-frame UV. Similarly to Ly$\alpha$, the components are mostly separated in the east-west direction and the western component extends somewhat towards the south. In case we interpret that the Ly$\alpha$ emission is indeed the combination of two LAEs for which we see the systemic redshifts in [CII] emission, we infer that the main component has a Ly$\alpha$ peak separation $\Delta v_{\rm Ly\alpha} = +213^{+19}_{-20}$ km s$^{-1}$ and that the redder component has $\Delta v_{\rm Ly\alpha} = +457^{+24}_{-19}$ km s$^{-1}$. This indicates that the two components have different HI column densities \citep[e.g.][]{Verhamme2006}. We discuss this interpretation in \S $\ref{sec:nature}$.

\section{Discussion} \label{sec:discussion}
The re-ionisation of the Universe is likely still ongoing at $z\approx6.5$, with a global neutral fraction of $\approx40$ \% \citep[e.g.][]{Naidu2019}. This could significantly impact the local UV background and therefore the HI structures in the nearby CGM of galaxies \citep[e.g.][]{MasRibas2017,Sadoun2017}, particularly for star-forming galaxies that are still in a local neutral bubble. Naively, one would expect that galaxies that reside in such a significantly more neutral environment have flatter Ly$\alpha$ SB profiles and observed Ly$\alpha$ lines with higher velocity shift (due to an increased importance of resonant scattering and the IGM damping wing), compared to post-re-ionisation LAEs \citep[e.g.][]{Dijkstra2007}. Are the Ly$\alpha$ properties of VR7 different from similar and lower-redshift galaxies? We now compare the observed spatial and spectral Ly$\alpha$ properties of VR7 to comparable galaxies at $z=5-6$ (i.e. just after re-ionisation) and other galaxies at similar redshift.

\subsection{Is Ly$\alpha$ emission more extended at $z>6$ than at later times?}
\begin{figure}
\includegraphics[width=8.7cm]{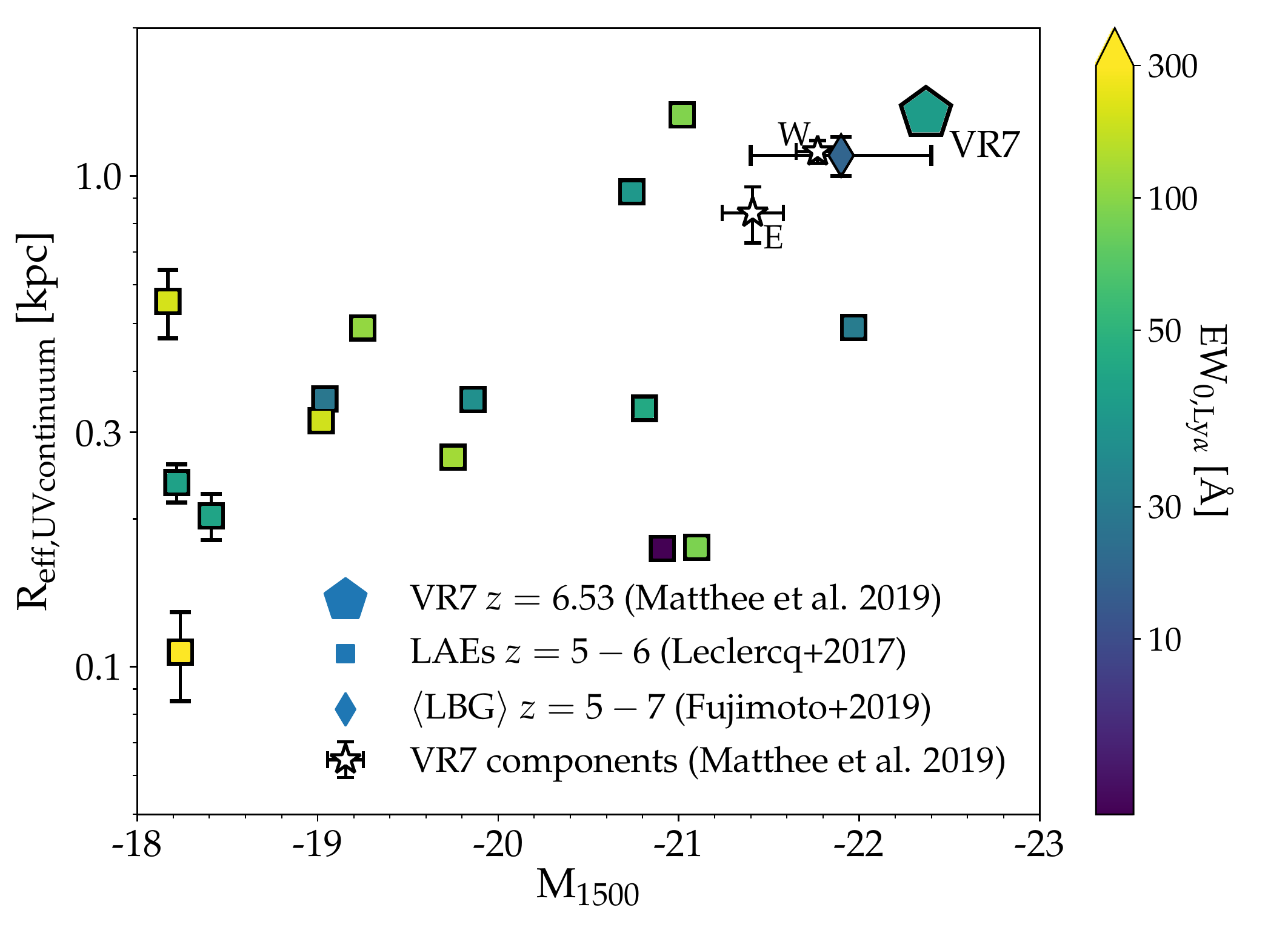}\\
\includegraphics[width=8.7cm]{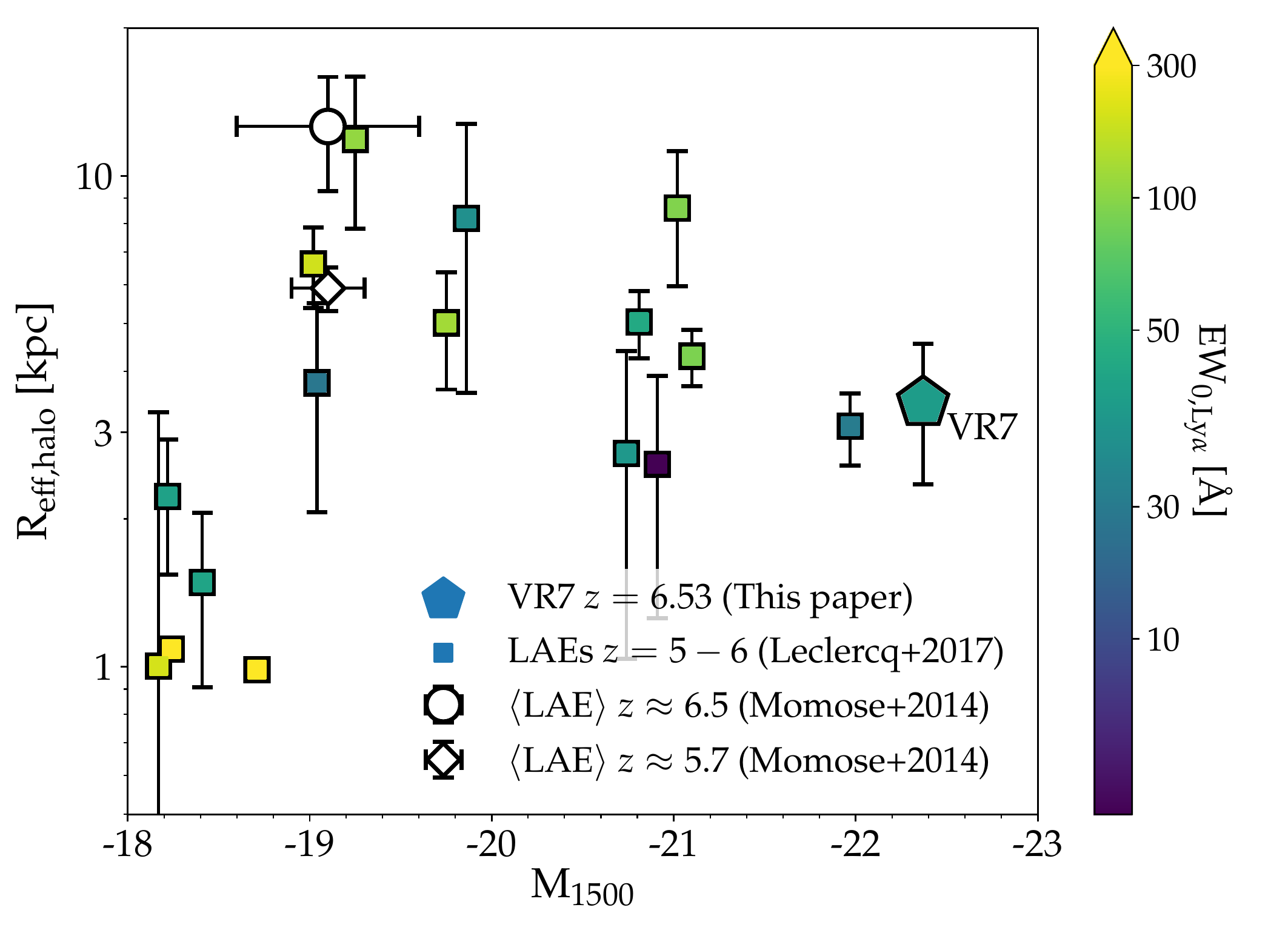}
    \caption{The dependence of UV (top panel) and Ly$\alpha$ (bottom panel) scale-length on the UV continuum luminosity for VR7 and a comparison sample of LAEs at $z=5-6$ from \citet{Leclercq2017}. We computed the median UV luminosity of the sample from \citet{Momose2014} using measurements of the same galaxy sample from \citet{Ono2010}. We also show the two individual UV components of VR7 as measured in \citet{Matthee2019}. The continuum scale length increases with luminosity and VR7 follows the trend of galaxies at $z=5-6$. There is no clear dependence between Ly$\alpha$ halo scale length and luminosity and VR7 has similar Ly$\alpha$ halo scale length as galaxies at LAEs $z=5-6$. }
\label{fig:extent_comparison} 
\end{figure}

In Fig. $\ref{fig:extent_comparison}$, we compare the UV continuum and Ly$\alpha$ halo scale lengths of VR7 to measurements of individual LAEs at $z=5-6$ by \cite{Leclercq2017} and stacked LAEs at $z=5.7-6.6$ by \cite{Momose2014}. We also compare the continuum scale length to a stack of UV-selected galaxies at $z=5-7$ by \cite{Fujimoto2019}. We note that we use our measurements based on a single UV component in VR7 to resemble the techniques in other works. 

As seen in Fig. $\ref{fig:extent_comparison}$, the rest-frame UV continuum scale length is observed to increase with luminosity and VR7 and its individual components follow this trend, with no indications of evolution with redshift. The Ly$\alpha$ halo scale length does not show a clear dependence on luminosity or redshift. While \cite{Momose2014} find a larger scale length at $z=6.6$ compared to $z=5.7$, the measured scale length at $z=6.6$ is similar to some individual LAEs at $z<6$ observed by \cite{Leclercq2017}. The halo scale length of VR7 is very comparable to the most luminous system in the sample from \cite{Leclercq2017}, but smaller than the scale length of typical LAEs at $z\approx6.5$. These results highlight that there is significant variation in halo scale lengths at fixed redshift and UV luminosity. There are no significant differences between the Ly$\alpha$ halo scale-lengths of VR7 and galaxies at $z<6$, which is expected if the HI column densities in the CGM are comparable \cite[e.g.][]{Sadoun2017}. Larger samples with better overlapping dynamic ranges are required both at $z>6$ and $z=5-6$ in order to identify more subtle, potentially luminosity dependent trends.

\subsection{Is there evolution of Ly$\alpha$ velocity offsets at $z>6$?}
Ly$\alpha$ observables are affected by gas on the interstellar medium (ISM), CGM, and IGM scales. Understanding the interplay of these scales is important, especially at higher redshifts where Ly$\alpha$ is used to put constraints on the epoch of reionisation. For example, the observed velocity offset between Ly$\alpha$ and the systemic redshift is an important ingredient in using the Ly$\alpha$-emitting fraction of high-redshift galaxies to measure the neutral fraction of the IGM \citep[e.g.][]{Mason2018}. However, if there are smaller velocity offsets at $z\approx7$ compared to $z\approx5$ due to an evolution in the ISM or CGM, the fraction of galaxies observable in Ly$\alpha$ emission will be lower at $z\approx7$ compare to $z\approx5$ \citep[e.g.][]{Choudhury2014}, even though there could be no difference in IGM properties. 

Therefore, evolution in the intrinsic velocity offset is degenerate to an evolution of the neutral fraction in the IGM, which also decreases the observed fraction of Ly$\alpha$ emitters \citep[e.g.][]{Pentericci2016}. Additionally, it has been argued \citep[e.g.][]{Stark2017,Mason2018} that, due to outflows, the Ly$\alpha$ velocity offset is larger in more luminous systems, facilitating their Ly$\alpha$ observability in the epoch of reionisation. However, the interpretation of large {\it observed} velocity offsets may be challenging in the epoch of re-ionisation. The IGM damping wing could cut-off a significant fraction of the flux on the blue parts of the line if a galaxy is surrounded by significant amounts of hydrogen \citep[e.g.][]{MiraldaEscude1998,Dijkstra2007,Laursen2011,Smith2019}. This will result in a large observed velocity offset. This could well be the case in the galaxy B14-65666 at $z\approx7$ which has a Ly$\alpha$ velocity offset of $\approx +800$ km s$^{-1}$ \citep{Hashimoto2018Dragons} and an accordingly low Ly$\alpha$ EW. 

\begin{figure}
\includegraphics[width=8.6cm]{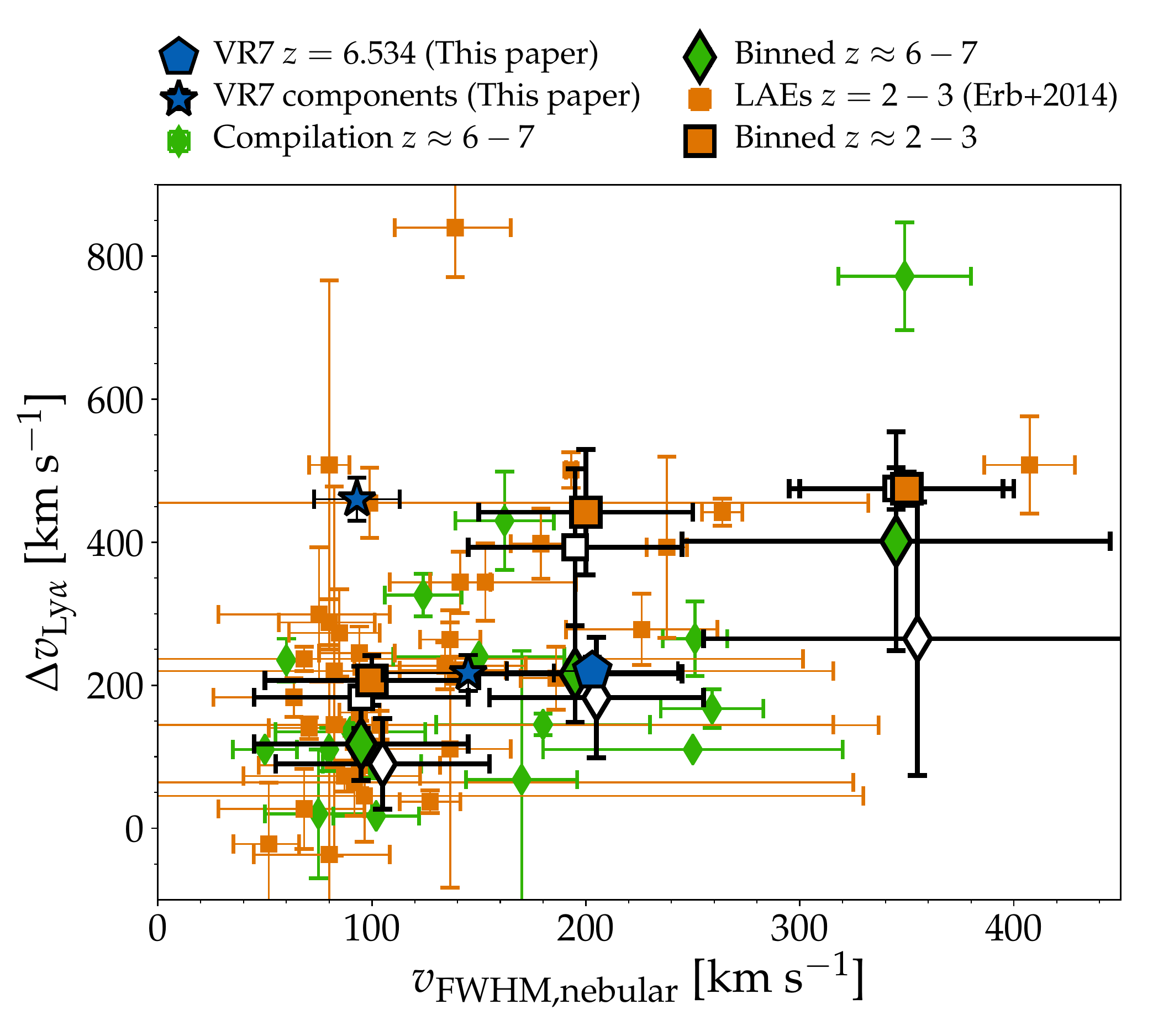}\\
\caption{The velocity offset between Ly$\alpha$ and the systemic redshift, $\Delta v_{\rm Ly\alpha}$, versus the nebular line-width (traced by [C{\sc ii}] or [OIII]). We show measurements for VR7 (total as blue pentagon; individual components as blue stars), a compilation of galaxies at $z\approx6-7$ (green diamonds) and Ly$\alpha$ emitters at $z\approx2-3$ from \citet{Erb2014} (orange squares). Larger symbols show {\it mean} velocity offsets in bins of line-width, where the error bars show the error on the mean. {\it Median} velocity offsets are shown with larger white-filled symbols. Galaxies at high-redshift have lower {\it observed} velocity offsets than galaxies at $z\approx2-3$. This indicates that the ISM in galaxies at $z\approx6-7$ is more ionised and that there is little additional absorption from circum/inter-galactic gas compared to galaxies at lower redshift.  \label{fig:delta_v_correlations}} 
\end{figure}

As the number of Ly$\alpha$ resonant scattering events is highly sensitive to the HI column density \citep{Neufeld1991}, a smaller Ly$\alpha$ velocity shift is found in case the ISM is more ionised \citep{Barnes2011} or more porous \citep[e.g.][]{GronkeDijkstra2016a,Smith2019}. Early results from ALMA measurements of the [OIII]$_{\rm 88 \mu m}$/[CII]$_{\rm 158 \mu m}$ ratio in a few galaxies indeed indicate a highly ionised ISM in galaxies at $z\sim7$ \citep{Inoue2016,Carniani2017,Hashimoto2018Dragons}. In Fig. $\ref{fig:delta_v_correlations}$ we compare the velocity shift between the observed peak of the Ly$\alpha$ emission to the systemic redshift ($\Delta v_{\rm Ly\alpha}$) with the width of nebular, non-resonant emission lines. We compare VR7 to other UV and Ly$\alpha$ selected galaxies at $z\sim6-7$ (for which [CII] is used as nebular line) and to LAEs at $z=2-3$ (for which [OIII]$_{5007 {\AA}}$ is used; \citealt{Erb2014}). VR7 has rather typical line-width and velocity shift for the $z\approx6-7$ population, although there may be significant dispersion for individual components. Both the mean and median observed velocity shifts in galaxies at $z\sim6-7$ are smaller than in Ly$\alpha$ emitters at $z\sim2-3$ at fixed nebular line-width, particularly at $\sigma\approx200$ km s$^{-1}$, although we note there is large scatter. This similarly points towards a more ionised ISM in observed LAEs at $z\sim6-7$.\footnote{We also note that at fixed properties, Ly$\alpha$ line-widths at $z\sim6-7$ are narrower than at $z\sim2-3$ \citep{Sobral2018}.} 

The implication of lower observed velocity shifts at high-redshift is that the majority of galaxies at $z\sim6-7$ that are observed in Ly$\alpha$ emission do not experience a strong additional HI damping wing compared to galaxies at $z\sim2-3$. In combination with the Ly$\alpha$ surface brightness profile this suggests that there is no detectable neutral hydrogen enhancement in both down-the-barrel and transverse direction. This agrees with the scenario that these galaxies reside in relatively large ionised regions. 

\subsection{How well can [CII] be used to measure Ly$\alpha$ velocity offsets?}\label{sec:issues_CII}
One potential caveat for using [CII] as a proxy to obtain the true systemic redshift is that [CII] has a lower ionisation energy than hydrogen. Because of this, the observed [CII] emission can trace a variety of gas phases. In local low-metallicity star-forming galaxies [CII] is predominantly comes from neutral gas \citep{Cormier2019}, but [CII] may also originate from HII regions or trace molecular gas \citep{Zanella2018}. If there are velocity differences between the regions which emit [CII] and the region that emit Ly$\alpha$ (i.e. HII regions), [CII] would not be an accurate means of systemic redshift. Furthermore, [CII] emission could trace neutral gas on which Ly$\alpha$ photons resonantly scatter. On the other hand, observations of both [CII] and [OIII] lines in galaxies at $z\sim7$ find consistent redshifts and line-widths \citep{Hashimoto2018Dragons,Walter2018}, suggesting that there is no problem in using [CII] for measuring Ly$\alpha$ velocity offsets. Moreover, observations of [CII] and molecular lines in sub-millimetre galaxies at $z\sim5$ also report consistent systemic redshifts \citep[e.g.][]{Riechers2014,Jimenez2018}.

Another caveat in interpreting Ly$\alpha$ velocity shifts is whether Ly$\alpha$ spectra are observed at the same spatial locations as [CII]. In VR7, we identify a (slight) spatial gradient in the peak velocities of both lines (\S $\ref{sec:profile_variations}$). Typically Ly$\alpha$ spectra of galaxies in the epoch of re-ionisation are obtained through narrow slits and can thus suffer from such intrinsic line profile variations. In VR7, [CII] is more extended than the UV continuum emission and resembles the Ly$\alpha$ extent. Interestingly, in the luminous LAE `Himiko' at $z=6.59$, the Ly$\alpha$ peak SB coincides with the peak in [CII] SB (and not with one of the UV components; \citealt{Carniani2018}). Furthermore, the [CII] emission that coincides with the region where Ly$\alpha$ peaks is emitted at a different velocity from the regions where [CII] and the rest-frame UV overlap. However, such offsets between Ly$\alpha$ and the UV are not seen in other luminous LAEs like CR7 \citep[e.g.][]{Sobral2019} and VR7 \citep[e.g.][]{Matthee2017ALMA,Matthee2019}. Future joint spatially resolved, high resolution spectroscopy of H$\alpha$ and Ly$\alpha$ emission would relieve these caveats, but current instruments can already be used to explore whether important spatial variations and offsets between Ly$\alpha$, UV and [CII] are also present in other galaxies.

\subsection{On the origin of variations in the the Ly$\alpha$ profile in VR7} \label{sec:nature}
In \S $\ref{sec:linevariations}$ we showed that the Ly$\alpha$ profile in VR7 has significant spatial variations. Do these variations originate in the emitting gas distribution and kinematics (for example two merging galaxies) or are they mostly driven by differences in the scattering medium, or both? 

The fact that the positions of the Ly$\alpha$ components (main peak and `bump') differ and resemble the positions of the UV and [CII] components (Fig. $\ref{fig:Lya_2comp}$) indicates towards a scenario where the variations are mainly driven by the emitting gas distribution. The dynamical information of the components seen in [CII] emission is not fully lost in the observed Ly$\alpha$ line profile, indicating the Ly$\alpha$ line profile is determined to large extent by processes relatively close to the galaxy. 

The grey lines in Fig. $\ref{fig:shellmodel}$ show the Ly$\alpha$ profile where the velocity axis is shifted as $v_{\rm new} = (v_{\rm obs}-260$ km s$^{-1})/2.6$ km s$^{-1}$. The values of 260 km s$^{-1}$ and 2.6 are motivated as they best match the peak separation in the [CII] line. It is surprising that this simple re-scaled version of the Ly$\alpha$ profile is able to resemble the red part of the [CII] profile well, particularly in the east. This requires a relatively specific distribution of expansion velocities and hydrogen column densities, in particular to have similar relative fluxes of the peaks. Within our limited current knowledge of the ISM properties of VR7, variations in the HI column densities and kinematics are plausible. The eastern part of the galaxy has a lower [CII]-UV ratio compared to the western part \citep{Matthee2019}, which could indicate low gas density \citep{Ferrara2019}. Future resolved multiple-line characterisation of the ISM properties in VR7 are required to address this question in more detail.

\section{Conclusions} \label{sec:conclusions} 
In this paper, we have presented spatially resolved Ly$\alpha$ spectroscopy of VR7 \citep{Matthee2017SPEC} with VLT/MUSE. VR7 is a luminous star-forming galaxy at $z=6.53$ that is resolved in two components in the UV and [CII]. Ly$\alpha$ emission is detected with an integrated S/N$\approx40$ and well resolved spatially and spectrally. We showed that the MUSE data is not deep and/or wide enough to accurately quantify the over-density of LAEs around VR7, only ruling out over-densities of a factor $>10$. We connected the Ly$\alpha$ line profile of VR7 to the velocity properties of the ISM (as traced by the [CII] emission observed by ALMA) for the first time at the epoch of re-ionisation. We searched for specific imprints of incomplete re-ionisation on the observed Ly$\alpha$ properties of VR7, such as a strongly broadened and/or redshifted and/or largely extended Ly$\alpha$ line, but find no significant trend.
Our main results are the following:

\begin{itemize}
\item Ly$\alpha$ emission (with a line-width FWHM=$374^{+21}_{-23}$km s$^{-1}$) in VR7 is more extended than the UV continuum, with a scale length r$_{\rm eff, Ly\alpha} = 2.05\pm0.16$ kpc compared to r$_{\rm eff, UV} = 1.34\pm0.06$ kpc (\S $\ref{sec:SB_Lya}$). The scale length of [CII] emission is similar to Ly$\alpha$ with r$_{\rm eff, [CII]} = 2.14^{+0.24}_{-0.22}$ kpc (\S $\ref{sec:SB_HSTALMA}$). Combining the Ly$\alpha$ with the UV data, we de-convolve the Ly$\alpha$ emission in a UV-like component and an extended halo-component with scale length r$_{\rm eff, Ly\alpha, halo} = 3.45^{+1.08}_{-0.87}$ kpc (\S $\ref{sec:SB_2comp}$). The halo scale length is comparable to UV-bright LAEs at $z=5-6$ observed by MUSE, but smaller (by a factor $\approx3.5$) than the stacked halo scale length of fainter LAEs at $z=6.6$  (Fig. $\ref{fig:extent_comparison}$). 

\item We identify spatial variations in the Ly$\alpha$ line profile (\S $\ref{sec:profile_variations}$). There is a tentative weak gradient in the peak velocity, redshifted by $\approx40$ km s$^{-1}$ in the south-western side of the galaxy compared to the north-east. We identify a redshifted bump in the eastern part of the Ly$\alpha$ line, which is redshifted by $\approx230$ km s$^{-1}$ with respect to the main Ly$\alpha$ peak (Fig. $\ref{fig:PV}$). According to the shell-model, the bump could correspond to back-scattering photons, but we find that the relative positions of the main Ly$\alpha$ component and the bump resemble those of components identified in {\it HST} rest-frame UV data. These components have a projected separation of $\approx2$kpc (Fig. $\ref{fig:Lya_2comp}$).

\item The main peak of the Ly$\alpha$ line is offset by $+213^{+19}_{-20}$ km s$^{-1}$ compared to the main peak of the [CII] line, but the spatial variations seen in the Ly$\alpha$ profile qualitatively resemble the variations in the [CII] line (Fig. $\ref{fig:PV}$). [CII] displays a similar, weak, peak gradient and a second eastern component. However, the [CII] line-width is narrower than Ly$\alpha$ and the velocity separation between the [CII] peak and the [CII] bump is smaller (by a factor $\approx2$). While a single shell model can accurately fit the Ly$\alpha$ profiles in different locations in the galaxy, its fitted intrinsic lines are somewhat redshifted and broader compared to the observed [CII] widths (Fig. $\ref{fig:shellmodel}$). The spatial and spectral resemblance of [CII] and Ly$\alpha$ indicates that the total observed Ly$\alpha$ emission in VR7 likely originates from (at least) two spatially and spectrally distinct regions. As the velocity offsets between the Ly$\alpha$ line and [CII] vary between the components, different HI column densities are plausibly present.

\item Using a literature compilation, we find that the velocity offsets between Ly$\alpha$ and the systemic are smaller at $z\approx6-7$ than found in LAEs at $z\approx2-3$ at fixed nebular line-width (Fig. $\ref{fig:delta_v_correlations}$). This indicates that the ISM in observed higher redshift galaxies is more ionised than at $z\approx2-3$. The observed Ly$\alpha$ photons from galaxies at $z\approx6-7$ do not experience a strong additional HI damping wing compared to galaxies at $z\sim2-3$. Therefore, these galaxies (including VR7) likely reside in relatively large ionised bubbles.
\end{itemize}

Our work reveals that constraints on the epoch of re-ionisation relying on Ly$\alpha$ observables need to take the potential evolution in the neutral hydrogen properties of the ISM and CGM into account. This will likely loosen the existing constraints significantly. The solution to break the major degeneracies is to explore the evolution of other Ly$\alpha$ observables such as the SB profile and the spectral properties as well as their correlation with different measurements. In this work we show the potential of such efforts on an individual galaxy. This can be put on solid, statistical grounds with larger future programs.

\section*{Acknowledgments}
We thank the referee for their suggestions and constructive comments that helped to improve the presentation of our results. Based on observations obtained with the Very Large Telescope, program 99.A-0462. Based on observations made with the NASA/ESA Hubble Space Telescope, obtained at the Space Telescope Science Institute, which is operated by the Association of Universities for Research in Astronomy, Inc., under NASA contract NAS 5-26555. These observations are associated with program \#14699. This paper makes use of the following ALMA data: ADS/JAO.ALMA\#2017.1.01451.S. ALMA is a partnership of ESO (representing its member states), NSF (USA) and NINS (Japan), together with NRC (Canada) and NSC and ASIAA (Taiwan) and KASI (Republic of Korea), in cooperation with the Republic of Chile. The Joint ALMA Observatory is operated by ESO, AUI/NRAO and NAOJ. MG acknowledges support from NASA grant NNX17AK58G. GP and SC gratefully acknowledge support from Swiss National Science Foundation grant PP00P2\textunderscore163824. BD acknowledges financial support from the National Science Foundation, grant number 1716907. We have benefited greatly from the public available programming language {\sc Python}, including the {\sc numpy, matplotlib, scipy} \citep{Scipy,Hunter2007,Numpy} and {\sc astropy} \citep{Astropy} packages, the astronomical imaging tools {\sc SExtractor, Swarp} and {\sc Scamp} \citep{Bertin1996,Bertin2006,Bertin2010} and the {\sc Topcat} analysis tool \citep{Topcat}.




\bibliographystyle{mnras}

\bibliography{bib_LAEevo.bib}




\appendix

\section{Consistency check HST based continuum - MUSE continuum} \label{app:UVcont} 
In the main text we subtract the UV continuum flux in the Ly$\alpha$ narrow-band using a model based on {\it HST}/WFC3 data. In this model, VR7 is described by a combination of two exponential profiles separated by $0.35''$ with scale lengths of $0.84$ and $1.12$ kpc and contributing 36 and 64 \% to the total flux, respectively (see \citealt{Matthee2019} for details). We use {\sc Imfit} \citep{Erwin2015} to create a model image that is convolved with the MUSE-PSF and normalise the flux based on extrapolating the UV luminosity and slope at $\lambda_{0}=1500$ {\AA} to the $\lambda_0\approx1230$ {\AA}. Once the model is convolved with the MUSE-PSF, it is well-fitted by a single exponential profile with r$_{\rm eff}=1.34$ kpc (after again accounting for the PSF).

\begin{figure*}
\includegraphics[width=15.6cm]{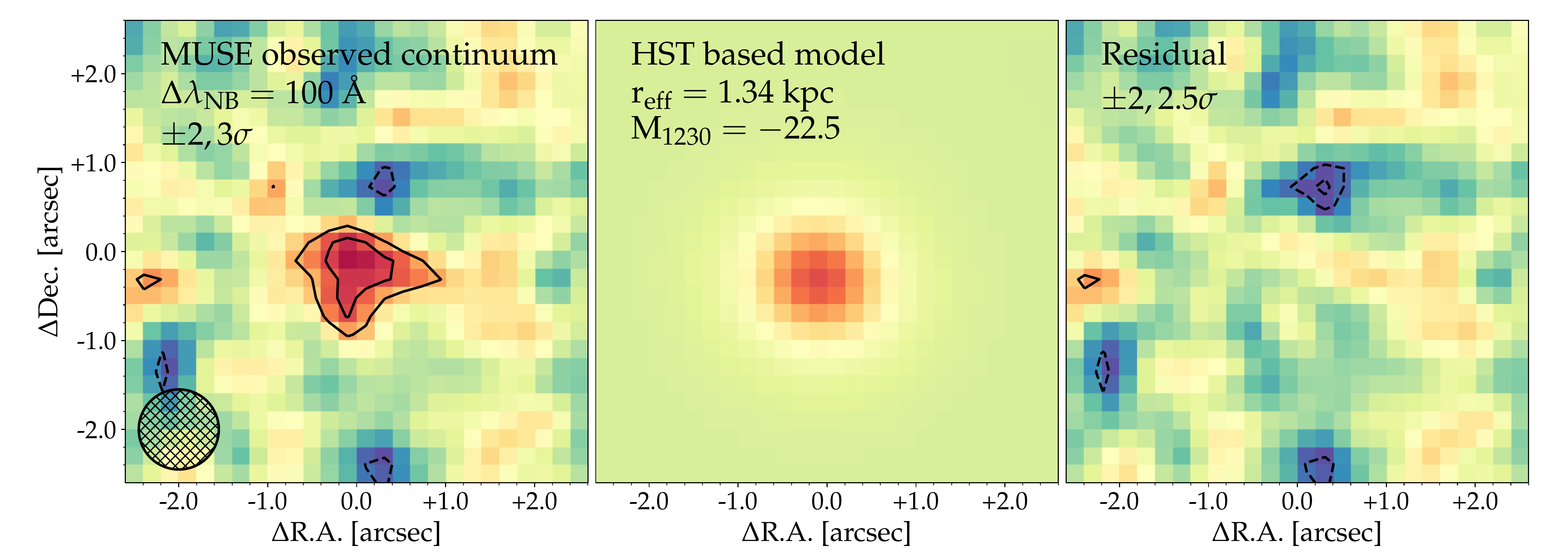}
\caption{Validation of our continuum subtraction model. The left panel shows the continuum detected in a collapsed image of $\lambda_{obs}=920-930$ nm in the MUSE cube. The middle panel shows the UV continuum model based on convolving the {\it HST} morphology with the MUSE PSF. The right panel shows the residual image. }
\label{fig:MUSE_CONT}
\end{figure*}

We perform a consistency check by comparing the UV continuum at $\lambda_0\approx1230$ {\AA} in our MUSE data to the prediction based on our model. The result is shown in Fig. $\ref{fig:MUSE_CONT}$. The left panel shows the detection of the UV continuum in our MUSE data (with S/N$\approx4$), while the middle panel shows the convolved {\it HST} based model. The right panel shows that no significant residuals are seen (except for the amplification of a negative noise peak already present in the data), providing a rough validation of our model.

\bsp	
\label{lastpage}
\end{document}